\documentclass[a4paper,11pt]{article}
\usepackage{jcappub} 
\usepackage{lineno}
\usepackage{natbib}
\usepackage[T1]{fontenc}
\usepackage{xcolor}
\usepackage{cleveref}
\usepackage{textcomp}
\usepackage{comment}
\DeclareUnicodeCharacter{2299}{\textcircled{\scriptsize  \ }}

\newcommand{\thetav}{\boldsymbol{\theta}}
\newcommand\aj{\ref@jnl{AJ}}
\newcommand\psj{\ref@jnl{PSJ}}
\newcommand\araa{\ref@jnl{ARA\&A}}
\newcommand\apj{\ref@jnl{ApJ}}
\newcommand\apjl{\ref@jnl{ApJL}}     
\newcommand\apjs{\ref@jnl{ApJS}}
\newcommand\ao{\ref@jnl{ApOpt}}
\newcommand\apss{\ref@jnl{Ap\&SS}}
\newcommand\aap{\ref@jnl{A\&A}}
\newcommand\aapr{\ref@jnl{A\&A~Rv}}
\newcommand\aaps{\ref@jnl{A\&AS}}
\newcommand\azh{\ref@jnl{AZh}}
\newcommand\baas{\ref@jnl{BAAS}}
\newcommand\icarus{\ref@jnl{Icarus}}
\newcommand\jaavso{\ref@jnl{JAAVSO}}  
\newcommand\jrasc{\ref@jnl{JRASC}}
\newcommand\memras{\ref@jnl{MmRAS}}
\newcommand\mnras{\ref@jnl{MNRAS}}
\newcommand\pra{\ref@jnl{PhRvA}}
\newcommand\prb{\ref@jnl{PhRvB}}
\newcommand\prc{\ref@jnl{PhRvC}}
\newcommand\prd{\ref@jnl{PhRvD}}
\newcommand\pre{\ref@jnl{PhRvE}}
\newcommand\prl{\ref@jnl{PhRvL}}
\newcommand\pasp{\ref@jnl{PASP}}
\newcommand\pasj{\ref@jnl{PASJ}}
\newcommand\qjras{\ref@jnl{QJRAS}}
\newcommand\skytel{\ref@jnl{S\&T}}
\newcommand\solphys{\ref@jnl{SoPh}}
\newcommand\sovast{\ref@jnl{Soviet~Ast.}}
\newcommand\ssr{\ref@jnl{SSRv}}
\newcommand\zap{\ref@jnl{ZA}}
\newcommand\nat{\ref@jnl{Nature}}
\newcommand\iaucirc{\ref@jnl{IAUC}}
\newcommand\aplett{\ref@jnl{Astrophys.~Lett.}}
\newcommand\apspr{\ref@jnl{Astrophys.~Space~Phys.~Res.}}
\newcommand\bain{\ref@jnl{BAN}}
\newcommand\fcp{\ref@jnl{FCPh}}
\newcommand\gca{\ref@jnl{GeoCoA}}
\newcommand\grl{\ref@jnl{Geophys.~Res.~Lett.}}
\newcommand\jcp{\ref@jnl{JChPh}}
\newcommand\jgr{\ref@jnl{J.~Geophys.~Res.}}
\newcommand\jqsrt{\ref@jnl{JQSRT}}
\newcommand\memsai{\ref@jnl{MmSAI}}
\newcommand\nphysa{\ref@jnl{NuPhA}}
\newcommand\physrep{\ref@jnl{PhR}}
\newcommand\physscr{\ref@jnl{PhyS}}
\newcommand\planss{\ref@jnl{Planet.~Space~Sci.}}
\newcommand\procspie{\ref@jnl{Proc.~SPIE}}

\newcommand\actaa{\ref@jnl{AcA}}
\newcommand\caa{\ref@jnl{ChA\&A}}
\newcommand\cjaa{\ref@jnl{ChJA\&A}}
\newcommand\jcap{\ref@jnl{JCAP}}
\newcommand\na{\ref@jnl{NewA}}
\newcommand\nar{\ref@jnl{NewAR}}
\newcommand\pasa{\ref@jnl{PASA}}
\newcommand\rmxaa{\ref@jnl{RMxAA}}

\title{Primordial black holes versus their impersonators at gravitational wave observatories}

\author[a,b]{Andrea Begnoni}
\affiliation[a]{Dipartimento di Fisica Galileo Galilei, Università di Padova, I-35131 Padova, Italy}
\affiliation[b]{INFN Sezione di Padova, I-35131 Padova, Italy}
\emailAdd{andrea.begnoni@phd.unipd.it}

\author[c,d]{and Stefano Profumo}
\affiliation[c]{Santa Cruz Institute for Particle Physics,\\Santa Cruz, CA, 95064, USA}
\affiliation[d]{Department of Physics, University of California, Santa Cruz \\Santa Cruz, CA, 95064, USA}
\emailAdd{profumo@ucsc.edu}

\abstract{The detection of primordial black holes (PBHs) would mark a major breakthrough, with far-reaching implications for early universe cosmology, fundamental physics, and the nature of dark matter. Gravitational wave observations have recently emerged as a powerful tool to test the existence and properties of PBHs, as these objects leave distinctive imprints on the gravitational waveform. Notably, there are no known astrophysical processes that can form sub-solar mass black holes, making their discovery a compelling signal of new physics. 
In addition to PBHs, we consider other exotic compact object (ECO) candidates—such as strange quark stars and boson stars—which can produce similar gravitational signatures and potentially mimic PBHs. 
In this work, we employ the Fisher matrix formalism to explore a broad parameter space, including binary masses, spins, and different nuclear and quark matter equations of state. Our goal is to assess the ability of next-generation gravitational wave detectors—specifically Cosmic Explorer and the Einstein Telescope—to distinguish PBHs from ECOs, stellar BHs and neutron stars. We compute the maximum luminosity distances (LDs) at which confident ($\geq 3\sigma$) detections of sub-solar masses or tidal effects are possible, providing quantitative benchmarks for PBH identification or exclusion under various observational scenarios. 
Our results indicate that next-generation detectors will be capable of probing sub-solar mass PBHs out to cosmological distances of $z \sim 3$. For heavier objects with masses up to $M \lesssim 2 M_\odot$, we show that PBHs can be distinguished from neutron stars via their lack of tidal effects up to redshifts of $z \sim 0.2$.}

\begin{document}

\newcommand{\mBH}{m_{\text{BH}}}
\newcommand{\mBHseed}{\mBH^{\text{seed}}}
\newcommand{\mBHobs}{\mBH^{\text{obs}}}
\newcommand{\mBHobsi}{m_{\text{BH},i}^{\text{obs}}}
\newcommand{\mBHtheory}{\mBH^{\text{theory}}}
\newcommand{\zcoll}{z_{\text{coll}}}
\newcommand{\zvir}{z_{\text{vir}}}
\newcommand{\zobs}{z_{\text{obs}}}
\newcommand{\cross}{\sigma/m}
\newcommand{\msun}{M_{\odot}}
\newcommand{\tsal}{t_{\text{sal}}}
\newcommand{\trel}{t_{\text{rel}}}
\newcommand{\rhocrit}{\rho_{\text{crit}}}
\newcommand{\cmg}{\text{cm}^{2}\text{g}^{-1}}
\newcommand{\kms}{\text{km}~\text{s}^{-1}}
\newcommand{\angstrom}{\r{A}}
\newcommand\sbullet[1][.5]{\mathbin{\vcenter{\hbox{\scalebox{#1}{$\bullet$}}}}}
\newcommand{\chisq}{\chi^{2}}
\newcommand{\Vmax}{V_{\text{max}}}

\newcommand{\spr}[1]{{\color{red}\bf[SP:  {#1}]}}
\newcommand{\ab}[1]{{\color{orange}\bf[AB:  {#1}]}}

\maketitle
\flushbottom

\section{Introduction}



The discovery of new classes of compact objects beyond the standard astrophysical paradigm
would offer profound insights into early-universe cosmology and the nature of dark matter.
Among the most intriguing candidates are primordial black holes~\cite{Carr:1974nx,
Hawking:1971ei} and exotic compact objects~\cite{Giudice:2016zpa}, such as strange quark
stars (SQSs)~\cite{Witten:1984rs, Postnikov:2010yn} or boson stars~\cite{Colpi:1986ye,
Yoshida:1997qf}, which could exist in mass regimes inaccessible to standard stellar
evolution~\cite{Crescimbeni:2024tidal, DeLuca:2024, Crescimbeni:2024qrq}. In particular,
the detection of a \textit{sub-solar mass} compact object would serve as a compelling
signature of new physics, since no known astrophysical process is expected to produce black
holes or neutron stars (NSs) below the Chandrasekhar limit. Recent searches within the
LIGO-Virgo-KAGRA (LVK) data have already begun placing stringent constraints on such
populations. The observation of gravitational waves (GWs) from compact binary mergers
provides a unique avenue for probing these
objects~\cite{Nitz:2022ltl, Wang:2016ana, LVK:2022ydq, LIGOScientific:2021sio}.

To distinguish between various sub-solar candidates, one must identify the distinct imprints
they leave on the GW signal. A central role is played by the \textit{tidal deformability},
characterized by the quadrupolar Love number $k_2$ and the dimensionless parameter
\begin{equation}
    \Lambda = \frac{2}{3} k_2 \left( \frac{R}{M} \right)^5,
\end{equation}
which encodes the internal structure and composition of the companions during the late
inspiral phase~\cite{Postnikov:2010yn, Binnington:2009bb, Damour:2012yf}, for a body of
radius $R$ and mass $M$.

An especially interesting regime is that of {\it sub-TOV masses}, where the object mass
falls below the maximum mass supported by the neutron-star Tolman-Oppenheimer-Volkoff (TOV)
limit. For material objects in this regime (NSs, SQSs, or BSs), the compactness $C = M/R$
is relatively low and the tidal deformability $\Lambda$ can be
large~\cite{Postnikov:2010yn, Perot:2023ghi}. By contrast, PBHs are vacuum solutions of GR
and have strictly vanishing tidal deformability, $k_2 = 0$ and $\Lambda = 0$ in classical
GR~\cite{Binnington:2009bb, Gurlebeck:2015xpa, DeLuca:2024}. The detection of a
sub-solar-mass black hole with $\Lambda \simeq 0$ would thus strongly indicate a primordial
origin, whereas a sub-solar-mass compact object with $\Lambda \gg 0$ would point to a
material configuration (NS or ECO)~\cite{Crescimbeni:2024tidal, Crescimbeni:2024qrq,
Sennett:2017etc}.

From a waveform modeling perspective, tidal contributions enter the GW phase at
$5^{\text{th}}$ post-Newtonian (PN) order and predominantly affect the late inspiral.
Their impact is enhanced for low-mass, low-compactness objects, but remains challenging to
measure even with third-generation (3G) detectors, owing to the high PN order and
correlations with other parameters~\cite{Vines:2011cz, Postnikov:2010yn,
Crescimbeni:2024tidal}. Accurate modeling of tidal terms is therefore crucial to avoid
biases in inferred masses, radii, and object
classification~\cite{Vines:2011cz, Gamba:2020ljo, Chatziioannou:2020pqz,
Crescimbeni:2024qrq}.

Upcoming observing runs of LIGO--Virgo--KAGRA and, in particular, future 3G observatories
such as Cosmic Explorer (CE)~\cite{Evans:2021gyd, Reitze:2019iox} and the Einstein
Telescope (ET)~\cite{Punturo:2010zz, ET:2019Magg} will dramatically improve sensitivity to
both sub-solar-mass systems and tidal signatures~\cite{pacilio_ranking_2022, Abac:2025BB}.
This raises a concrete question: {\it to what redshift can 3G detectors identify
sub-solar-mass compact objects, and within what distance can they distinguish PBHs from
material counterparts via tidal effects?} Answering these questions requires combining
realistic assumptions about detector networks and waveform models with physically motivated
mass and tidal-deformability ranges for NSs, SQSs, BSs, and PBHs.

In this work we address this problem using the \texttt{GWJulia} code~\cite{Begnoni:2025oyd}
and the Fisher information matrix (FIM)
formalism~\cite{Finn:1992wt, Cutler:1994ys, Vallisneri:2007ev}. The FIM has been widely
used for GW forecasts, including for 3G
detectors~\cite{Rodriguez:2013mla, Iacovelli:2022For, Dupletsa:2022scg,
Borhanian:2020GWB, Borhanian:2022czq, deSouza:2023DALI, Li:2021Tidofm, Abac:2025BB}, and
provides reliable estimates in the high-SNR
regime~\cite{Dupletsa:2022scg, Begnoni:2025oyd}. We use it here to quantify: (i) the
{\it mass-based identification horizon} for sub-solar companions, and (ii) the
{\it tidal-discrimination horizon} at which PBHs and other ECOs can be distinguished from
standard NSs.

The paper is organized as follows. In Sec.~\ref{sec:max_dL} we define our Fisher-matrix
setup and the criterion for mass-based sub-solar identification. In
Sec.~\ref{sec:compactobjects} we summarize the compact-object models considered in this
work, i.e., NSs, SQSs, BSs, and PBHs, focusing on their expected mass ranges and tidal
deformabilities. Section~\ref{sec:setup} details the detector network, waveform choices,
and EOS models. The main results are presented in Sec.~\ref{sec:results}, and we conclude
in Sec.~\ref{sec:conclusions}.

\vspace{0.5em}

\section{\label{sec:max_dL} Maximum luminosity distance}

Our first goal is to quantify, for a given compact-binary configuration in which one object is sub-solar, the maximum luminosity distance at which the mass of the sub-solar companion can be identified at the $3\sigma$ level. 
We do so using the Fisher information matrix, defined as~\cite{Finn:1992wt, Cutler:1994ys, Vallisneri:2007ev}
\begin{equation}\label{eq:FIM}
    \Gamma_{ij} \equiv 
    - \left\langle 
      \frac{\partial^2 \log \mathcal{L}(d | \boldsymbol{\theta})}
           {\partial \theta^i \partial \theta^j}
      \right\rangle_n ,
\end{equation}
where $\log \mathcal{L}(d|\boldsymbol{\theta})$ is the log-likelihood of the data stream $d$
given parameters $\boldsymbol{\theta}$, and the average $\langle \dots \rangle_n$ is taken over
noise realizations, see, \cref{app} for more details on the formalism. The parameter vector includes intrinsic and extrinsic quantities,
\begin{equation}
    \boldsymbol{\theta} =
    \left(
      \mathcal{M}_{\rm c}^{\rm det}, q, \chi_1, \chi_2, d_{\rm L}, \theta, \phi, 
      \iota, \psi, t_{\rm coal}, \Phi_{\rm coal}, \Lambda_1, \Lambda_2
    \right),
\end{equation}
where $\mathcal{M}_{\rm c}^{\rm det}$ is the chirp mass at detector, $q = m_2/m_1$ the mass ratio, $\chi_{1,2}$ the
dimensionless component spins, $d_{\rm L}$ the luminosity distance, $(\theta,\phi)$ the sky
position, $\iota$ the inclination, $\psi$ the polarization angle, and $t_{\rm coal}$,
$\Phi_{\rm coal}$ the time and phase at coalescence. For material objects (NSs or ECOs),
$\Lambda_{1,2}$ denote the tidal deformabilities; for PBHs we set $\Lambda = 0$. In this work we will consider non-precessing waveforms, therefore $\chi_i = \chi_{{\rm z}, i}$.

In the high-SNR regime, the likelihood of the event $\mathcal{L}(d\mid\thetav)$, converges to a Gaussian. Therefore, the covariance matrix of parameter estimates $\Sigma_{ij}$ is approximated by $\Sigma_{ij} \simeq (\Gamma^{-1})_{ij}$, and, for instance, the $1\sigma$ uncertainty on the mass ratio $q$ is given by $\sigma(q) = \sqrt{\Sigma_{qq}}$~\cite{Vallisneri:2007ev}. 

Since we want to obtain contraints on the lighter companion in the merger $m_2$ we need to perform a conversion of the FIM. The conversion is $(\mathcal{M}_{\rm c}^{\rm det}, q)\xrightarrow{} (m_1, m_2)$, which is performed assuming a cosmology, Planck-18~\cite{Planck:2018vyg} in our case. This is needed because the chirp mass at the source $\mathcal{M}_{\rm c}^{\rm source}$ is linked to the chirp mass measured at the detector with 
\begin{equation}
    \mathcal{M}_{\rm c}^{\rm source} = \mathcal{M}_{\rm c}^{\rm det} /(1+z)\,.
\end{equation} 
This, united with the definition of chirp mass at the source, allows us to perform the change of variables in the FIM.
Our primary diagnostic for sub-solar identification is the luminosity distance $d_{L,\,3\sigma}$
satisfying
\begin{equation}\label{eq:subsolar}
    m_2 + 3 \sigma(m_2)\vert_{ d_L = d_{L,\,3\sigma}} < 1\,M_\odot\,.
\end{equation}
therefore, we define $d_{L,\,3\sigma}$ as the maximum luminosity distance at which a sub-solar mass object can be confidently identified at the $3\sigma$ level. Our primary objective is to analyze how different configurations of intrinsic parameters (masses, spins, and EOS-dependent tidal deformabilities $\Lambda_{1,2}$) impact this detection horizon. However, the extrinsic parameters $\vec{\theta}_{\rm ext} = (\theta, \phi, \iota, \psi, t_{\rm coal}, \Phi_{\rm coal})$ can significantly influence the estimation of $d_{L,\,3\sigma}$. Ideally, one would marginalize the results over these variables to isolate the effects of the intrinsic physics; however, a full marginalization procedure is computationally prohibitive. 
To balance physical rigor with computational feasibility, we adopt a sampling approach: for each unique intrinsic configuration, we generate a catalog of 20 binaries with extrinsic parameters $(\theta, \phi, \iota, \psi, t_{\rm coal}, \Phi_{\rm coal})$ drawn from their respective astrophysical priors \cite{KAGRA:2021vkt, Begnoni:2025oyd}. This allows us to provide results that are representative of the expected observational population while remaining within manageable numerical limits.
Then, for each of the 20 realizations identified with parameters $\thetav^i$, we determine, via a bisection search on $d_{\rm L}$, the maximum LD $d_{L,\,3\sigma}^i$, using Eq. \eqref{eq:subsolar}. We then take the {\it median} over this catalog as our measure of the maximum LD $d_{L,\,3\sigma}$ (or equivalently, redshift) of that intrinsic configuration.
This median-based approach has several advantages. It averages over the strong variation in network response with sky position and the binary inclination and it yields a more robust measure. Moreover, it alleviates possible shortcomings of the FIM framework, as we discuss later.

In addition to mass-based identification, we also use the FIM to forecast the precision on tidal parameters and to determine the maximum redshift at which nonzero $\Lambda$ can be measured with $3\sigma$ significance, thereby excluding the PBH (or other ECO) hypothesis. The same catalog-based procedure is used, but with the relevant criterion applied to $\Lambda_2$, as detailed in Sec.~\ref{sec:results}.

While the FIM is a powerful tool, its convergence to the true posterior distribution is subject to several important caveats. GW likelihoods are frequently multimodal and can exhibit significant non-Gaussian tails, even at high SNRs. Parameters such as spins and tidal deformabilities are particularly susceptible to this behavior. For instance, low sensitivity to spin can cause the FIM to predict errors that exceed physical boundaries (e.g., $|\chi| > 1$). While some effects can be mitigated by enforcing prior bounds, such as the approach in \cite{Dupletsa:2024gfl}, others, like the strong degeneracy between individual tidal deformabilities, remain inherently non-Gaussian. Numerical stability also poses a significant hurdle. Inverting the FIM can be ill-conditioned, a problem exacerbated by the additional parameters introduced in tidal models \cite{Vallisneri:2007ev}. Similarly, face-on events produce nearly circular polarization, inducing near-total degeneracies between inclination, distance, phase, and polarization angle. Because the inversion process is sensitive to even a single poorly constrained parameter, a nearly singular matrix (characterized by near-zero eigenvalues) can lead to significant numerical errors. Despite these limitations, the FIM remains a standard when analyzing large event populations, e.g., \cite{Abac:2025BB}. Although individual low-SNR or degenerate events may yield biased results, the collective parameters inferred from a large catalog tend to converge toward the true distribution obtained via full Bayesian estimation \cite{Begnoni:2025mtz}. 

\vspace{0.5em}

\section{Compact-object models and tidal properties}
\label{sec:compactobjects}

In this section we summarize the compact-object families considered in our analysis and the aspects that are directly relevant for our Fisher forecasts. Our goal is 
to identify the characteristic mass ranges, compactness, and tidal deformabilities that determine how easily different objects can mimic or be distinguished from primordial black holes in 3G observations.

We consider four representative classes: (i) neutron stars modeled with the AP3 EOS, (ii) strange quark stars described by a self-bound quark-matter EOS, (iii) phenomenological boson stars whose tidal deformability is enhanced relative to AP3, and (iv) primordial black holes, which provide a null hypothesis with $\Lambda = 0$. For the material objects (NSs, SQSs, BSs), the key quantities are the allowed mass range and the mass--$\Lambda$ relation; for PBHs, the crucial features are the absence of tidal deformability and the freedom to populate the sub-solar mass range.

\subsection{Strange quark stars}
\label{sec:SQS}

Strange quark stars are hypothetical compact stars composed of deconfined up, down,
and strange quarks, motivated by the Bodmer--Witten hypothesis that strange quark matter
may represent the true ground state of hadronic matter~\cite{Witten:1984rs, Bodmer:1971we}.
Their macroscopic properties are governed by the EOS of strange quark
matter, which determines the mass--radius relation and thus the tidal deformability.

For our purposes, the key aspects of SQSs are:
\begin{itemize}
    \item \textbf{Mass range.} For viable EOSs compatible with massive pulsars such as
    PSR~J0740+6620 ($M \simeq 2.1\,M_\odot$)~\cite{Cromartie:2019kug}, SQSs can reach maximum
    masses $M_{\max} \sim 2\,M_\odot$, similar to neutron stars. Unlike NSs, however, self-bound
    SQS configurations can in principle extend to significantly lower masses, potentially
    well below $1\,M_\odot$ for sufficiently small bag constants or appropriate microphysics
    (see, e.g.,~\cite{Zdunik:2000xx, Benvenuto1998}).
    \item \textbf{Tidal deformability.} For a given mass, SQSs are typically more compact than NSs
    and therefore have smaller tidal deformabilities $\Lambda$. Representative models yield
    $\Lambda_{1.4}$ values spanning from $\mathcal{O}(10^2)$ to several $10^2$--$10^3$, depending
    on the bag constant, color-superconducting gap, and additional QCD corrections~\cite{Zhou:2017pha,Zhiqiang:2021}.
    This implies that, in the sub-TOV mass regime, SQSs can still exhibit sizable tidal
    signatures, but generically somewhat smaller than those of NSs at the same mass.
\end{itemize}

In our analysis we do not attempt to resolve the full microphysical parameter space of SQS
EOSs. Instead, we adopt a representative SQS model, denoted SQM3 \cite{Prakash:1995uw}, that is consistent with
current mass and tidal constraints and yields a mass--$\Lambda$ relation intermediate between
soft and stiff hadronic EOSs. This choice captures the qualitative behavior of SQS
tidal deformabilities in the sub-TOV regime and allows us to assess how easily SQSs could
mimic PBHs in 3G observations.

Spin properties of SQSs are less critical for our purposes. Existing modeling indicates that
rapidly rotating SQSs with $\chi \sim 0.5$--$0.7$ are possible for realistic EOSs before
reaching the mass-shedding limit~\cite{Bhattacharyya:2016nhb}. In the present work, spins
enter only through the adopted spin ranges in our waveform modeling, as summarized in
Sec.~\ref{sec:setup}.

\vspace{0.5em}

\subsection{Boson stars}
\label{sec:BS}

Boson stars are self-gravitating configurations of complex scalar or vector fields that
are stabilized by the balance between gravity and the field’s gradient and self-interaction
energies~\cite{Yoshida:1997qf, Schunck:1999zu}. They provide a well-studied example of
horizonless ECOs that can mimic black holes or neutron stars in mass and compactness, but
with distinct tidal properties.

The aspects of BSs that are directly relevant to our analysis are:
\begin{itemize}
    \item \textbf{Mass scaling.} For a free scalar field of mass $m_b$, the maximum BS mass
    scales as $M_{\max} \sim M_{\rm Pl}^2/m_b$~\cite{Kaup:1968zz}, while strong quartic
    self-interactions modify this to $M_{\max} \sim \sqrt{\lambda} M_{\rm Pl}^3 / m_b^2$
    in the appropriate regime~\cite{Colpi:1986ye}. This allows BSs with masses in the
    stellar range for a broad class of particle-physics models.
    \item \textbf{Compactness and tidal deformability.} Depending on $m_b$ and the self-interaction
    potential, BSs can range from relatively diffuse to highly compact. In general, lighter
    bosons and strong self-interactions favor more extended configurations with large
    tidal deformabilities $\Lambda$, whereas heavier bosons or certain solitonic potentials
    can produce very compact BSs with much smaller $\Lambda$~\cite{Chavanis:2011zi, Kleihaus:2011sx}.
\end{itemize}

Given this diversity, a fully microphysical exploration of BS parameter space would be far
beyond the scope of this work. Instead, we adopt a phenomenological BS model, denoted BS5,
whose mass--$\Lambda$ relation is chosen to follow that of the AP3 EOS but with tidal
deformabilities enhanced by a fixed factor (see Sec.~\ref{sec:setup} and \cref{app}).
This setup is deliberately optimistic: it maximizes the tidal imprint relative to NSs and PBHs,
and thus provides an optimistic estimate of how easily BSs could be distinguished from PBHs
with 3G detectors.

We do not model BS-specific merger or ringdown effects; instead, we treat BSs as material
objects characterized by an effective tidal deformability entering the inspiral waveform.

\vspace{0.5em}

\subsection{Lower mass limits of neutron stars}
\label{sec:NSlower}

Neutron stars are formed from the gravitational collapse of massive stellar cores and are
stabilized by neutron degeneracy pressure and strong interactions. For our purposes, the
key question is how low in mass a stable NS can be, and what tidal deformabilities are
expected in the low-mass regime relevant for sub-solar or sub-TOV systems.

Solutions of the TOV equations indicate that the absolute lower mass limit for cold, stable
NS configurations is of order $0.1$--$0.2\,M_\odot$ for realistic EOSs~\cite{Lattimer2004,
Lattimer:2000nx}. However, proto-NS evolution and supernova dynamics imply that NSs produced
in nature are unlikely to populate the very low-mass tail: the minimum {\it formation}
mass is typically at or above $\sim 1\,M_\odot$ due to thermal pressure and trapped neutrinos
in the proto-NS phase~\cite{Lattimer2004}. Observationally, well-measured NS masses lie
predominantly in the $\sim 1.1$--$2.1\,M_\odot$ range.

At fixed EOS, low-mass NSs are less compact and thus exhibit large tidal deformabilities.
For the AP3 EOS adopted in our analysis, $\Lambda$ reaches values of order $10^3$ for
$1.1\,M_\odot$ and decreases as the mass approaches the TOV maximum (see \cref{app}).
This strong mass dependence underlies the key idea of our tidal discrimination forecasts:
even modest fractional errors on $\Lambda$ at low mass translate into powerful constraints
on whether a sub-solar (or near-sub-solar) object can be consistent with a material NS.

\vspace{0.5em}

\subsection{Primordial black holes}
\label{sec:PBHs}

Primordial black holes are black holes formed in the early Universe from the collapse
of large density perturbations or other non-stellar processes such as phase transitions or
cosmic-string collapse~\cite{Carr:2023tpt, Green:2020jor}. Because their masses are set by
horizon-scale physics rather than stellar evolution, PBHs can populate a wide mass spectrum,
including the sub-solar regime inaccessible to standard astrophysical formation channels
(e.g.~\cite{Hawking:1971ei, Carr:1974nx, Carr:2016drx}).

For this work, the relevant PBH properties are \cite{franciolini_how_2022}:
\begin{itemize}
    \item \textbf{Mass range.} The mass spectrum of PBHs depends on the underlying formation
    scenario; however, sub-solar-mass PBHs with $M \lesssim 1\,M_\odot$ are a generic
    possibility in many models and would be extremely difficult to explain by stellar
    evolution alone~\cite{Suwa:2018yxd, Kacanja:2025}.
    \item \textbf{Tidal deformability.} Classical GR predicts that non-spinning black holes
    have exactly vanishing tidal Love numbers, $k_2 = 0$ and hence $\Lambda = 0$~\cite{Binnington:2009bb,
    Damour:2009vw, Poisson:2014cga}. This holds also for Kerr black holes at the PN level
    relevant for inspiral. As a result, PBHs provide a clean null hypothesis for tidal
    effects: any robust detection of nonzero $\Lambda$ for a sub-solar object would strongly
    disfavor a PBH interpretation.
\end{itemize}

PBH spins are more model dependent. PBHs forming during radiation domination are expected
to have low spins due to the near-sphericity of the initial perturbations~\cite{Chiba:2017rvs},
whereas formation in matter-dominated eras or subsequent accretion can generate higher
spins~\cite{Harada:2017fjm, DeLuca:2019buf}. In our forecasts, spins enter only through the
assumed spin ranges in the waveform models; we do not attempt to constrain detailed spin
distributions.

\section{Methods and Assumptions} \label{sec:setup}
The results are significantly dependent on the network of detectors considered. In this work, we consider a standard 3G network composed of 10 km arm-length triangular ET in Sardinia, plus two CE, one of 40 km arm-length in the US and one of 20 km arm-length also in the US. For more information on the detectors, see \cref{app}.
One of the improvements that the FIM analysis can bring is that we do not have to limit the analysis to a few sources, as described in \cref{sec:max_dL}. Therefore, in each cell of the figures shown, we plot the median maximum redshift calculated from a small catalog of 20 sources. These catalogs, obtained by uniformly sampling the extrinsic parameters (i.e., inclination, sky position, polarization angle, and phase and time of coalescence), were created to partially reduce the noise associated with a single realization and to obtain a more robust result, see \cref{sec:max_dL}. For each event, we obtain an estimate of the maximum redshift for a $3\sigma$ detection using a bisection method; subsequently, we take the median of the 20 redshift estimates. This leads to $\mathcal{O}(1-5\times10^5)$ FIM evaluation per plot.
The code used for the evaluation is \texttt{GWJulia}~\cite{Begnoni:2025oyd}, which enabled a very fast evaluation of the FIMs, making it feasible to run all the FIMs required for this work on a laptop.

We now proceed with setting up the Fisher Matrix evaluation, and an important question in each scenario is the waveform choice. We use the most advanced waveform at our disposal, given the physical constraints, since the computation time is not a significant issue for FIM analysis\footnote{\texttt{GWJulia} has also the possibility of adding higher order harmonics for the binary BH (BBH) case~\cite{Garcia-Quiros:2020XHM}; we reserve the option to add this in future work.}. In this work, we analyze different scenarios and summarize the information on the waveforms in \cref{tab:prim_sec_obj} and in \cref{tab:calibration}. For each waveform, in the \texttt{GWJulia} implementation, we use settings replicating the default options of \texttt{LALSuite}~\cite{lalsuite}, e.g., the frequency where to cut the waveform. 
The waveforms used in this work are \texttt{IMRPhenomXAS}~\cite{Pratten:2020XAS} for the BBH case, \texttt{IMRPhenomD\_NRTidal\_v2}~\cite{Husa:2015PhD, Khan:2015PhD, Dietrich:2019NRTidal} for the binary NS (BNS) case, \texttt{IMRPhenomNSBH}~\cite{Pannarale:2015NSBH, Thompson:2020PhenomNSBH} for the NSBH case, and \texttt{TaylorF2}~\cite{Pan:2007F2, Boyle:2009F2, Buonanno:2009F2} when working outside the calibration ranges of the previous waveforms.

The tidal deformability used for NS is obtained using the AP3 EOS~\cite{Akmal:1998cf}, with the TOV solver provided by \texttt{LALSuite}~\cite{lalsuite}. This EOS is compatible with the current constraints, in particular the one provided by GW170817~\cite{LIGOScientific:2018cki, pacilio_ranking_2022}. The EOS used for SQSs is the SQM3 \cite{Prakash:1995uw}.
For the BS tidal deformability, we consider a phenomenological approach, i.e., for each mass $M$ we do not fix the boson mass $m_b$. Instead, we consider a BS which is softer than a NS, choosing the tidal deformability to follow the AP3 EOS slope but with a five times larger magnitude. This allows us to place more competitive bounds on the tidal deformability of BS, since softer BS will have an even larger signature, without the need to focus on a single boson mass, $m_b$ or potential shape. For more information on the EOS used in this work, we refer to \cref{app}.

\begin{table}[h!]
    \centering
    \begin{tabular}{|c|c|c|c|}
    \hline
       object 1/ object 2  & PBH & NS &  ECO \\
       \hline   
        NS & \texttt{TaylorF2} & \texttt{IMRPhenomD\_Tidal}&  \texttt{TaylorF2}\\
        \hline      
        BH & \texttt{IMRPhenomXAS} & \texttt{IMRPhenomNSBH} & \texttt{TaylorF2}\\
        \hline
        ECO & / & /  & \texttt{TaylorF2} \\
        \hline
    \end{tabular}
    \caption{Table representing which waveform model is used in the different scenarios. The criterion for the choice is to use the most advanced waveform given the physical requirements. \texttt{IMRPhenomD\_Tidal} stands for \texttt{IMRPhenomD\_NRTidal\_v2}}
    \label{tab:prim_sec_obj}
\end{table}


\begin{table}[h!]    
    \centering
    \begin{tabular}{|c|c|c|c|c|c|}
    \hline
       waveform  & $1/q$ & $M_2\,[m_{\odot}]$  & $\chi_1$ & $\chi_2$ &$\Lambda$\\
       \hline   
        \texttt{IMRPhenomD\_NRTidal\_v2} & [1,3] & [1,3] & [-0.6,0.6] &[-0.6,0.6] &[0,5000]\\
        \hline
        \texttt{IMRPhenomXAS} & [1, 1000] & BBH & [-0.9,0.9] & [-0.9,0.9] & 0.\\
        \hline
        \texttt{IMRPhenomNSBH} & [1,15] & [1,3] & [-0.5,0.5] & 0. & [0,5000] \\
        \hline
        
    \end{tabular}
    \caption{Summary of the calibration regimes of validity of the different waveforms used. Note that the [1,1000] for \texttt{IMRPhenomXAS} is the calibration range claimed in~\cite{Pratten:2020XAS}}
    \label{tab:calibration}
\end{table}
\section{Results}
\label{sec:results}

\begin{figure}[h!]
    \centering
    \includegraphics[width=\linewidth]{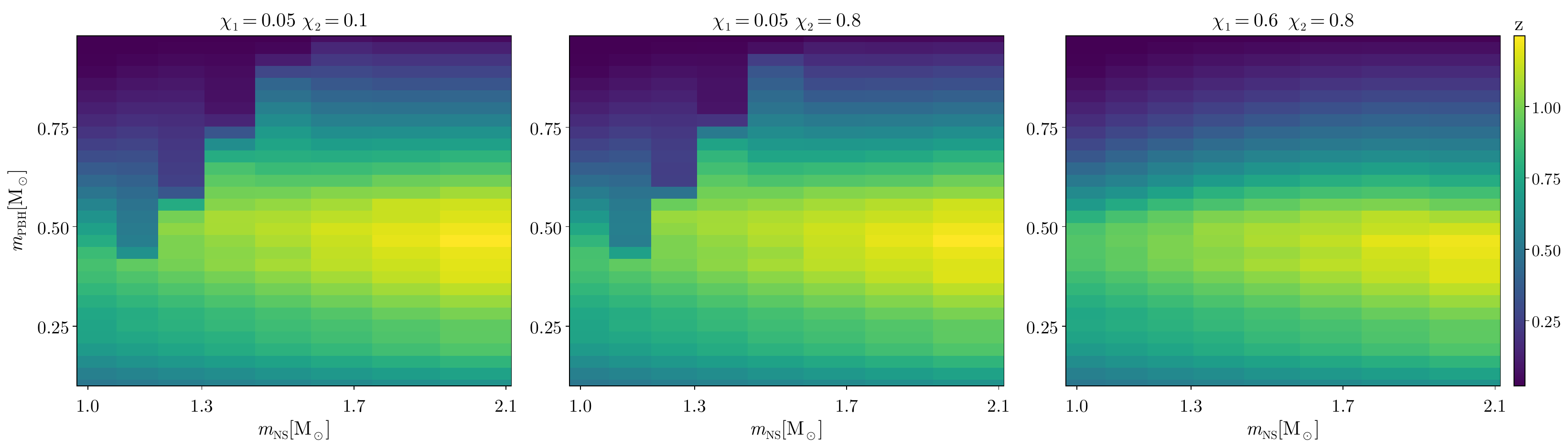}
    \caption{In these figures, we show the maximum redshift at which we can have a 3 sigma measurement of a sub-solar mass PBH, in the case where the merging partner is a {\it neutron star}. The first mass, on the $x$-axis, is a NS, while the PBH masses are on the $y$-axis. The different plots represent different spin configurations: low-spins (\textit{left}), high PBH spin (\textit{center}) and high aligned spins (\textit{right}).
    The waveform used is \texttt{TaylorF2} and each pixel represents the median redshift of 20 events. The largest redshift is reached for the heaviest NS considered ($M_{\rm NS} \sim 2.1 \, M_\odot$) in combination with a PBH of mass  $m_{\rm PBH} \sim 0.5\, M_\odot$.}
    \label{fig:max_dL_mass_subsolar_NS}
\end{figure}

In \cref{fig:max_dL_mass_subsolar_NS}, we show the maximum redshift at which a $3\sigma$ measure of a sub-solar mass PBH is possible in a merger with a neutron star (NS) of varying masses. Therefore we show the redshift $z_{3\sigma}$ such that 
\begin{equation}
    m_2 + 3 \sigma(m_2)\vert_{ z = z_{3\sigma}} < 1\,M_\odot\,.
\end{equation}
The waveform used is \texttt{TaylorF2}, which includes leading-order tidal effects but lacks merger and ringdown information—resulting in a conservative estimate of the sensitivity. This choice is made because none of the full inspiral-merger-ringdown waveforms respects physical constraints, i.e., the more massive body is a NS and the secondary body is a BH. The panels depict three spin configurations: low spins, high spin of the PBH and high aligned spins of both bodies. Spins have a modest impact on the inspiral phase, so we find very weak modifications in the different configurations. The optimal configuration is reached for $M_{\rm NS}\sim2.1\,M_\odot$ and $M_{\rm PBH}\sim0.5\,M_\odot$, with the median horizon redshift extending beyond $z\sim 1$. This highlights the importance of heavy neutron star companions in maximizing the reach of PBH identification. 

\begin{figure}[h!]
    \centering
    \includegraphics[width=\linewidth]{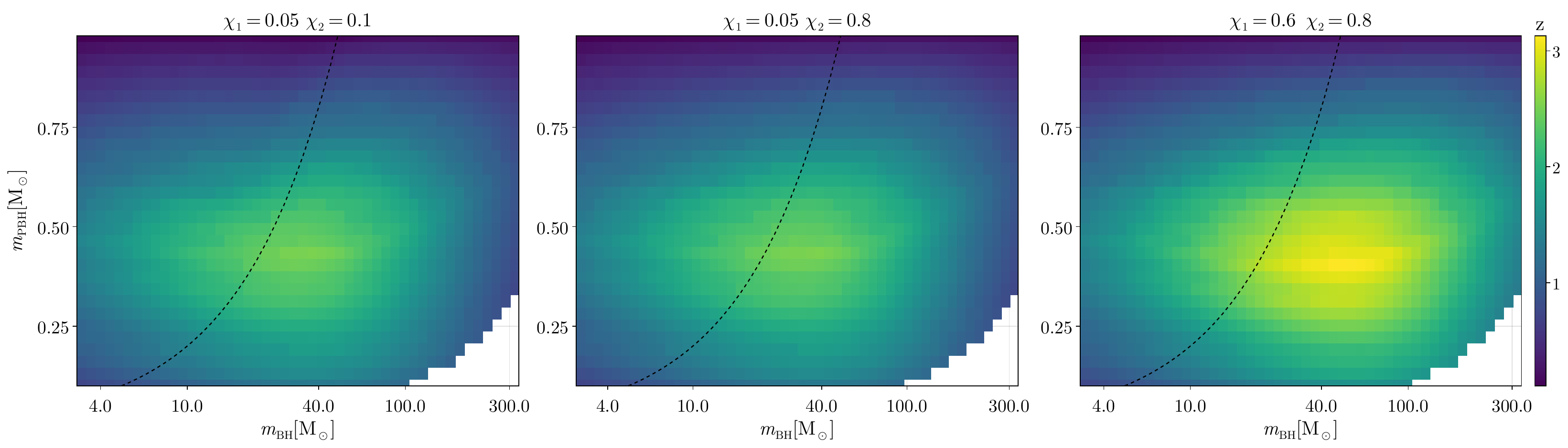}
    \caption{In these figures, we show the maximum redshift at which we can have a 3 sigma measurement of a sub-solar mass BH, in the case where the merging partner is a {\it black hole}. The first mass, on the $x$ axis, is a BH, while the PBH masses are on the $y axis$. The different plots represent different spin configurations: low-spins (\textit{left}), high PBH spin (\textit{center}) and high aligned spins (\textit{right}).
    The waveform used is \texttt{IMRPhenomXAS} and each pixel represents the median redshift of 20 events.nThe largest redshift is reached for $M_{\rm BH} \sim 40 \, M_\odot$) in combination with a PBH of mass  $m_{\rm PBH} \sim 0.4\, M_\odot$. The presence of high BH spin slightly enlarges the mass of the BH at which this maximum occurs. Moreover, both high-spin configurations lead to significantly larger median redshifts across all masses considered.
    The dashed lines indicate where the mass ratio $q=50$; above the lines, the waveform is highly reliable; in the white region, $q>1,000$ and the waveform lies outside the calibration range.}
    \label{fig:max_dL_mass_subsolar_BH}
\end{figure}

In \cref{fig:max_dL_mass_subsolar_BH}, we consider instead mergers of a sub-solar mass PBH with a stellar black hole companion and we still examine 3$\sigma$ sub-solar detection. This time, we use the \texttt{IMRPhenomXAS} waveform that allows us to account for high mass ratios. Three spin configurations are again analyzed, showing a more pronounced effect from the high spins alignment compared to the NS case, due to the inclusion of the merger in the evaluation. The maximum redshift is attained for BH mass $M_{\rm BH} \sim 40\,M_\odot$ and PBH masses around $M_{\rm PBH} \sim 0.4\,M_\odot$, reaching beyond $z \sim 3$. The left panel (low spins) shows the lowest redshift contours, while the right panel (high aligned spins) demonstrates the highest detection reach. The dashed contour denotes the threshold beyond which the mass ratio $q > 50$. The choice of such a threshold is strongly dependent on the goals of the analysis. E.g., in the context of Bayesian parameter estimations, similar thresholds are put in place; however, this choice is a complex topic, dependent on the waveform model, the spins, and total mass of the event~\cite{Rink:2024swg, Gadre:2022sed, Lam:2023oga, Ohme:2011zm}. In this work, we consider events inside the full calibration range of \texttt{IMRPhenomXAS}~\cite{Pratten:2020XAS}.
Beyond $q > 1000$, shown in white, the waveform lies outside the calibration range. This plot confirms that high-mass, high-spin BH companions enable the deepest reach for sub-solar PBHs.

\begin{figure}
    \centering
    \includegraphics[width=0.5\linewidth]{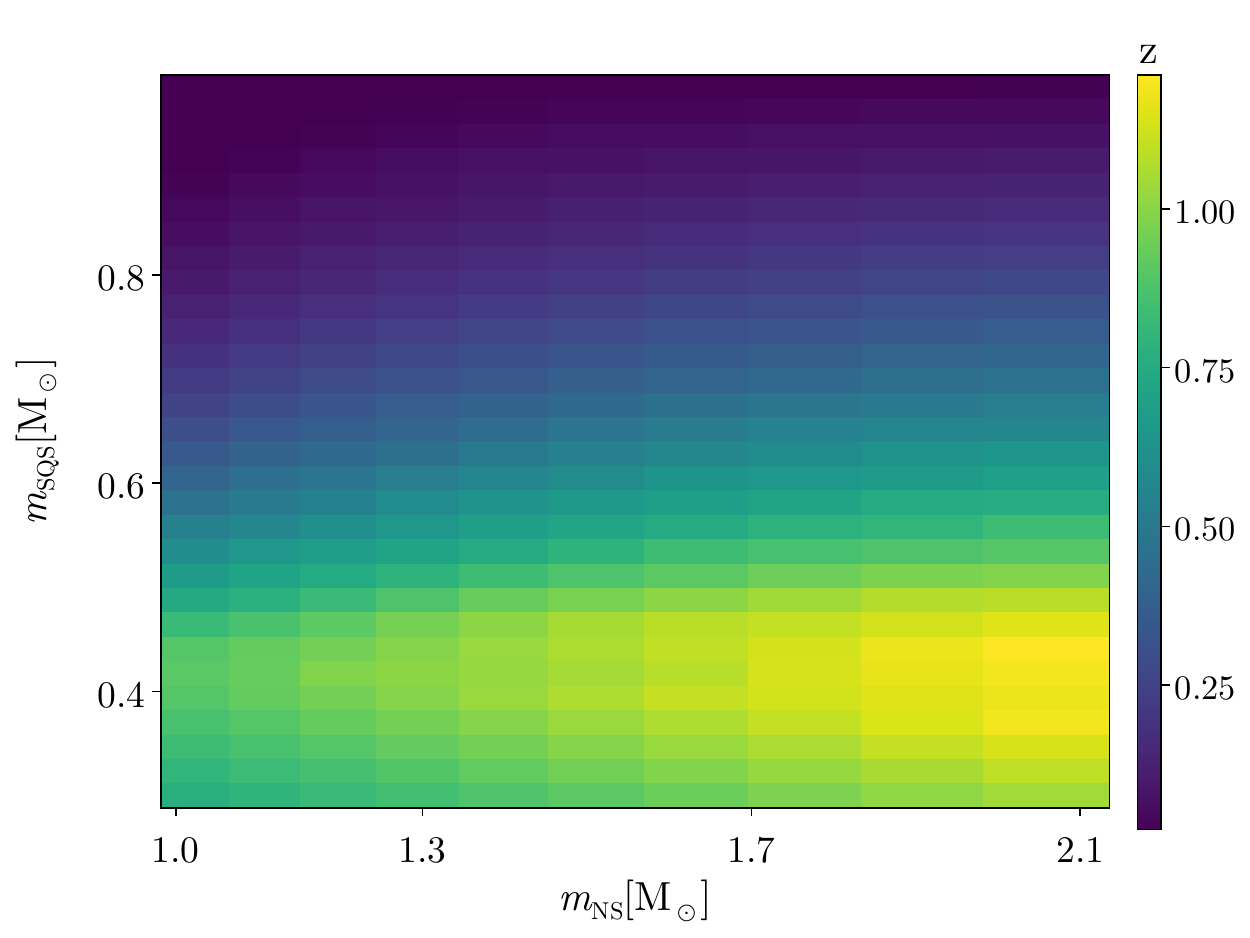}
    \includegraphics[width=0.49\linewidth]{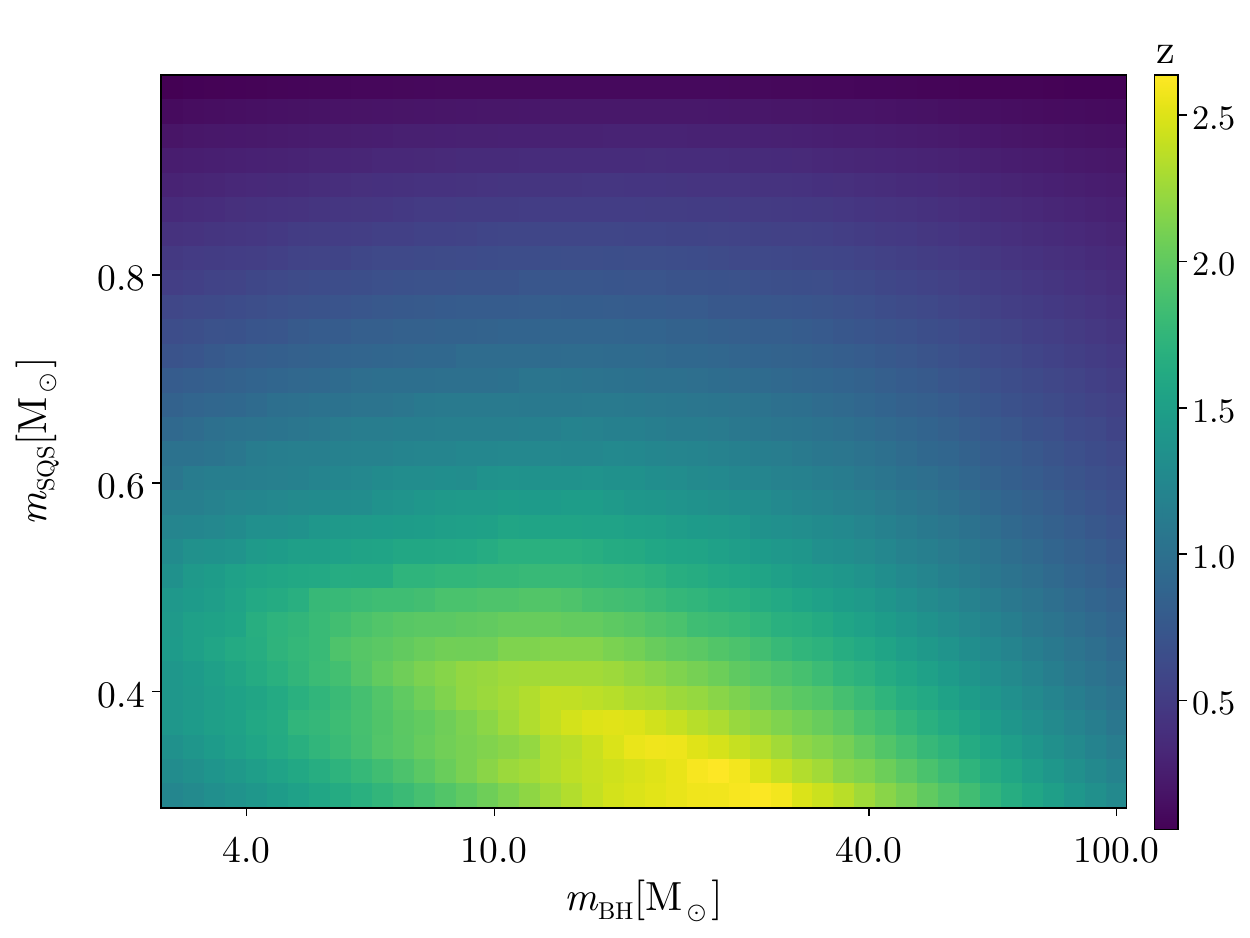}
    \caption{Maximum redshift at which a 3 sigma measurement of sub-solar mass is possible, for a SQM3 EOS. We consider the \texttt{TaylorF2} waveform model. Object 1 is a NS in the \textit{left} panel and a BH in the \textit{right} panel. These plots do not imply the ability to distinguish between a SQS and a PBH, as the tidal deformability at these redshifts is poorly constrained. The maximum redshift for a merger with the NS is found at the point $\sim(2.1, 0.4) M_\odot$, while for the merger with BH the point is at $\sim(30,0.3)M_\odot$.}
    \label{fig:max_dL_mass_subsolar_SQM}
\end{figure}


Fig.~\ref{fig:max_dL_mass_subsolar_SQM} displays the same detection calculation for a SQS as the low-mass companion, assuming the SQM3 EOS~\cite{Prakash:1995uw}. Results are shown for NS-SQS and BH-SQS systems and we evaluate only the low spin configuration ($\chi_1 = 0.05$, $\chi_2 = 0.1$), using the \texttt{TaylorF2} model. The trends are similar to those in the PBH case, with detection optimized for heavy companions and light SQSs (down to $0.3-0.4\,M_\odot$). However, these plots only reflect sub-solar mass detectability; distinguishing SQSs from PBHs requires measurement of nonzero tidal deformability, which we will show is inaccessible at such high redshifts. Thus, this figure sets bounds on the regime where exotic self-bound objects could be confused with PBHs. 

\begin{figure}
    \centering
    \includegraphics[width=0.5\linewidth]{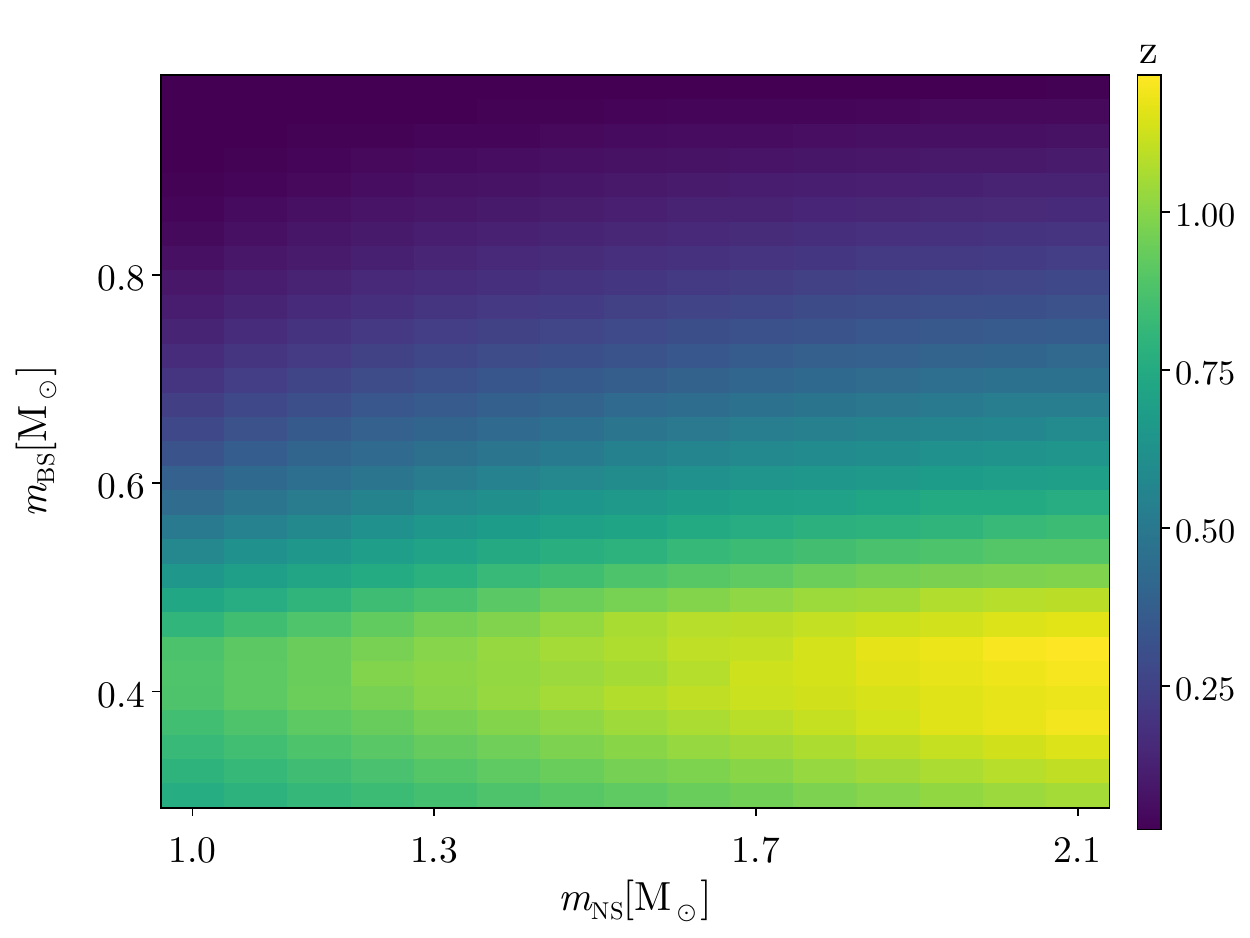}
    \includegraphics[width=0.49\linewidth]{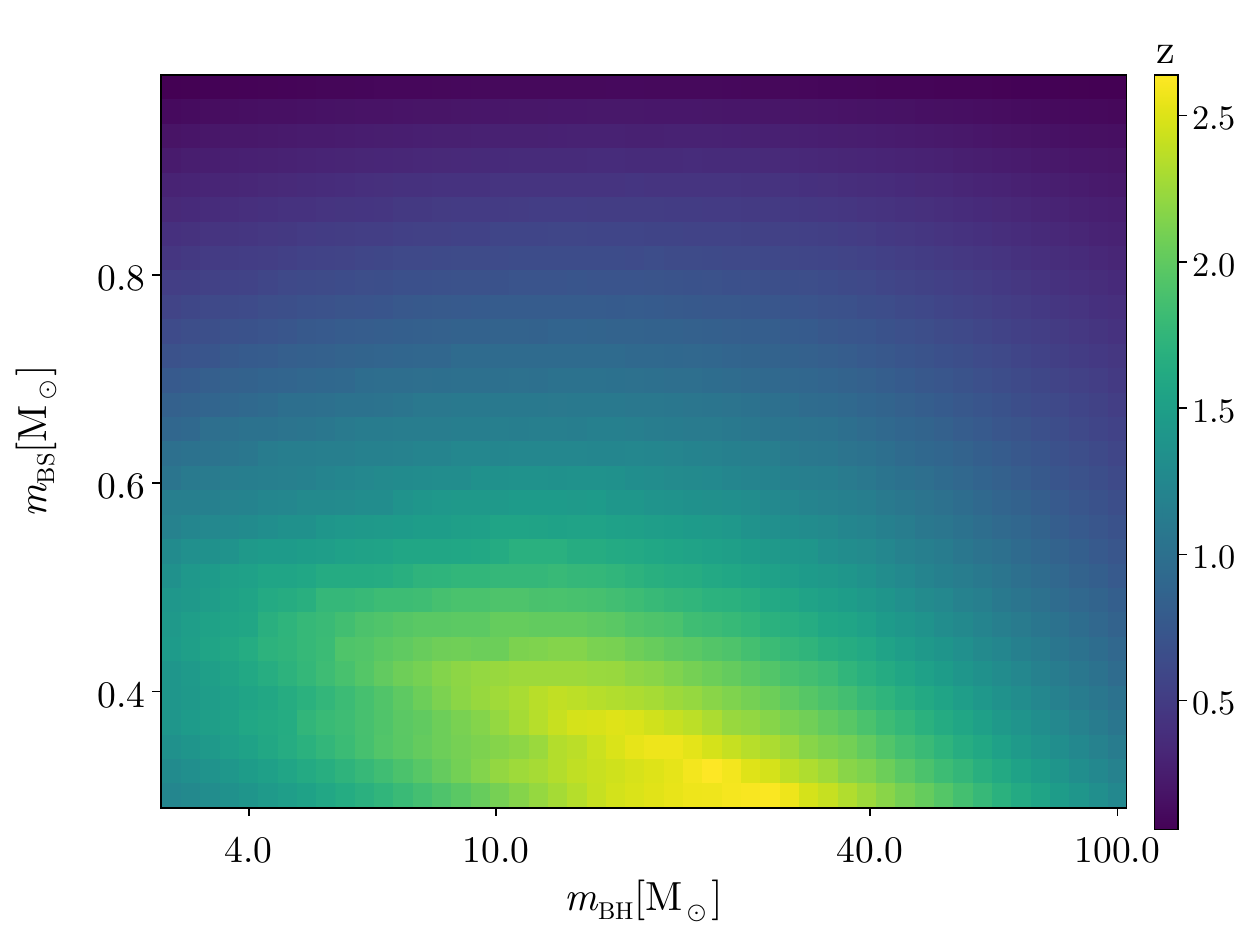}
    \caption{Maximum redshift at which we can have a 3 sigma measurement of sub-solar mass for a BS EOS. We consider an EOS producing a tidal deformability of 5 times the AP3 EOS, with the boson mass adapted accordingly. We considered the \texttt{TaylorF2} waveform model. Object 1 is a NS in the \textit{left} panel and a BH in the \textit{right} panel. Again, these plots do not imply distinguishability from a PBH via tidal deformability imprints. The maximum redshift for a merger with the NS is found at the point $\sim(2.1, 0.4) M_\odot$ while for the merger with BH the point is at $\sim(30,0.3)M_\odot$.}
    \label{fig:max_dL_mass_subsolar_BS_5}
\end{figure}

In \cref{fig:max_dL_mass_subsolar_BS_5}, we perform an analogous analysis for BS companions. We assume an EOS yielding a tidal deformability $\Lambda = 5 \times \Lambda_{\rm AP3}$ to amplify possible differences from PBHs. Also, here we evaluate only the low spin configuration ($\chi_1 = 0.05$, $\chi_2 = 0.1$) with the \texttt{TaylorF2} model. Even under this optimistic assumption, the tidal signature is too small to affect detectability at cosmological distances. More significant differences could be expected by using a full inspiral-merger-ringdown waveform, designed for BS~\cite{evstafyeva_gravitational-wave_2024}. The maximum redshift for detection again lies near the $(2.1, 0.4)\,M_\odot$ point for NS-BS systems and around $(30, 0.3)\,M_\odot$ for BH-BS systems, consistent with the SQS and PBH trends.

 Figs. 1--4 show that 3G detectors can identify sub-solar mass compact objects out to cosmological distances when the discrimination criterion is purely mass-based, i.e., when the measured secondary mass is found at more than $3\sigma$ to be below $1\,M_\odot$.  
This capability arises because the mass imprint enters at low  order in the inspiral phase and benefits directly from the enhanced low-frequency sensitivity of Cosmic Explorer and Einstein Telescope.  
For PBH-BH mergers (\cref{fig:max_dL_mass_subsolar_BH}), the maximum reach approaches $z \gtrsim 3$, and in favorable spin-mass configurations extends even farther.  
For PBH-NS systems (\cref{fig:max_dL_mass_subsolar_NS}), the reach typically lies in the $z \sim 1$-$2$ range, driven by the smaller total mass and correspondingly shorter effective inspiral.  
These results demonstrate that the {\it existence} of sub-solar mass compact objects can be probed deep into the high-redshift universe, independently of the nature of the object.

\begin{figure}
    \centering
    \includegraphics[width=0.5\linewidth]{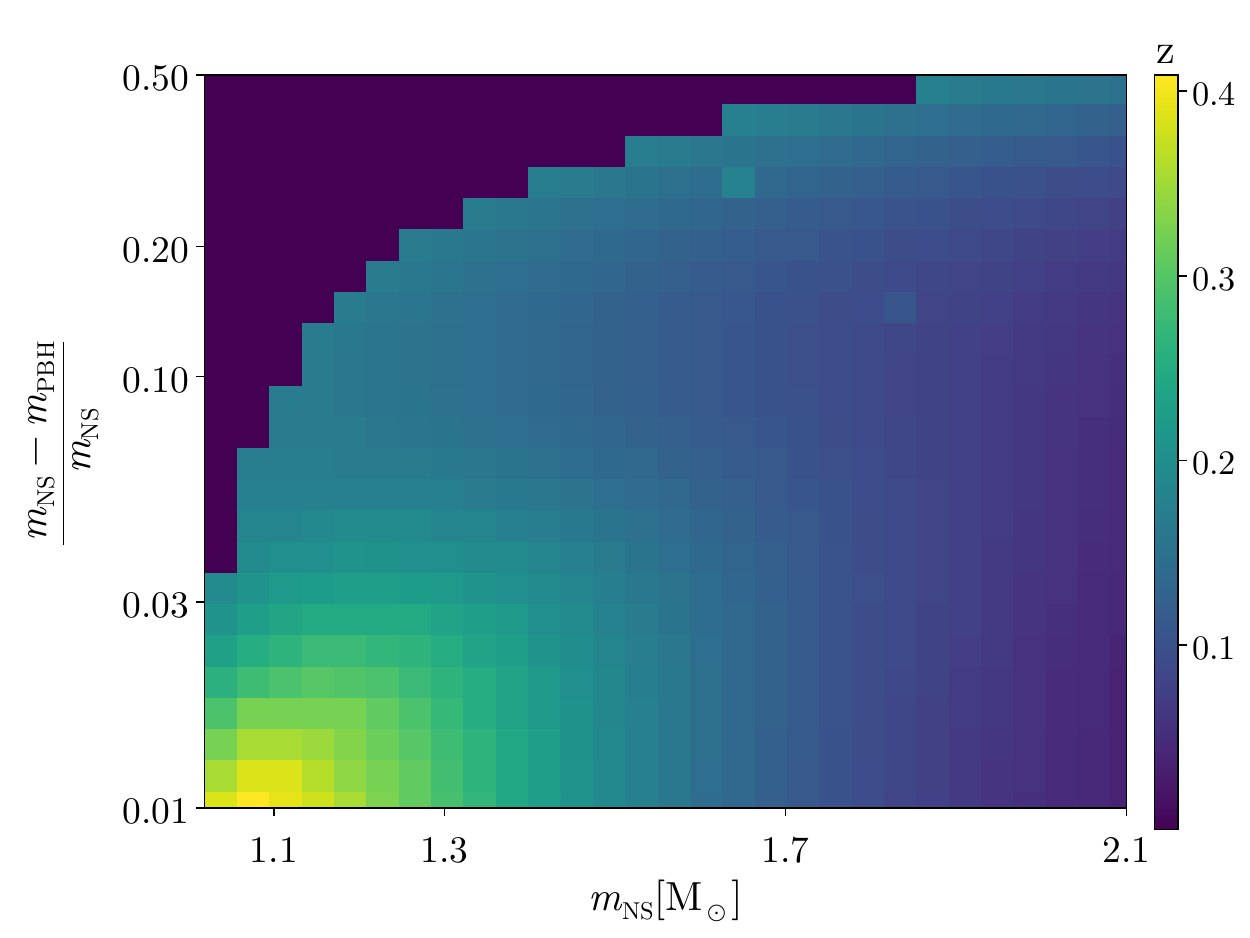}
     \includegraphics[width=0.49\linewidth]{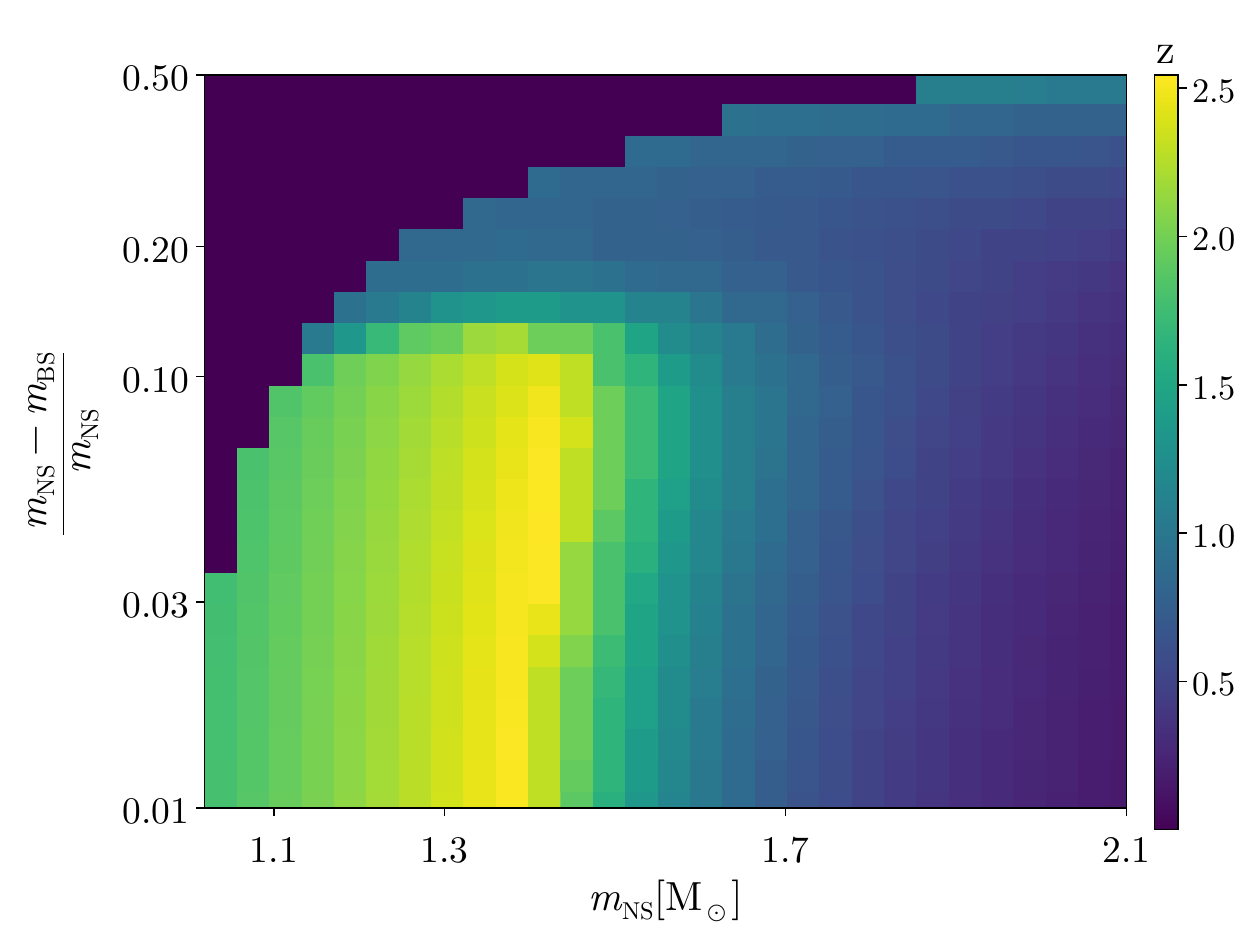}
     \caption{
    Maximum redshift at which we can have a $3\sigma$ exclusion of PBH (\textit{left}) or BS (\textit{right}) using tidal effects. The injected signal is a BNS merger, using \texttt{IMRPhenomD\_NRTidal\_v2}. We constrain the tidal parameter of the second NS $\Lambda_2$ after conditioning over the first tidal parameter $\Lambda_1$. We limit ourselves to objects larger than a solar mass and on the $y$ axis, we plot the relative difference of the two objects.
    The best constraints are for light NS of similar masses $\sim(1.1, 1.1) M_\odot$ in the PBH case, while for the BS case, it is at $\sim(1.4, 1.3)M_\odot$.   
    }
    \label{fig:max_dL_Lambda_NS}
\end{figure}

\begin{figure}
    \centering
    \includegraphics[width=0.5\linewidth]{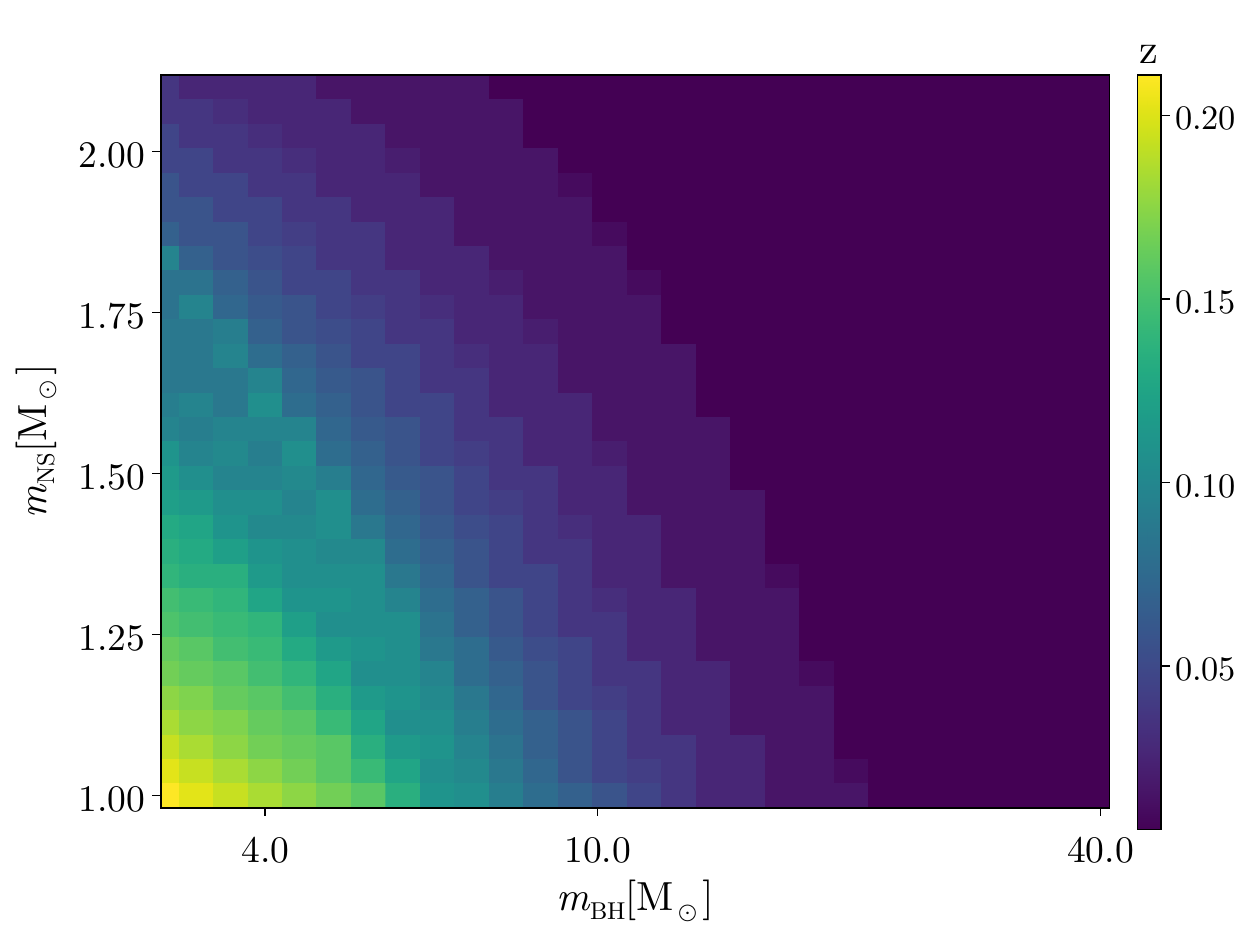}
     \includegraphics[width=0.49\linewidth]{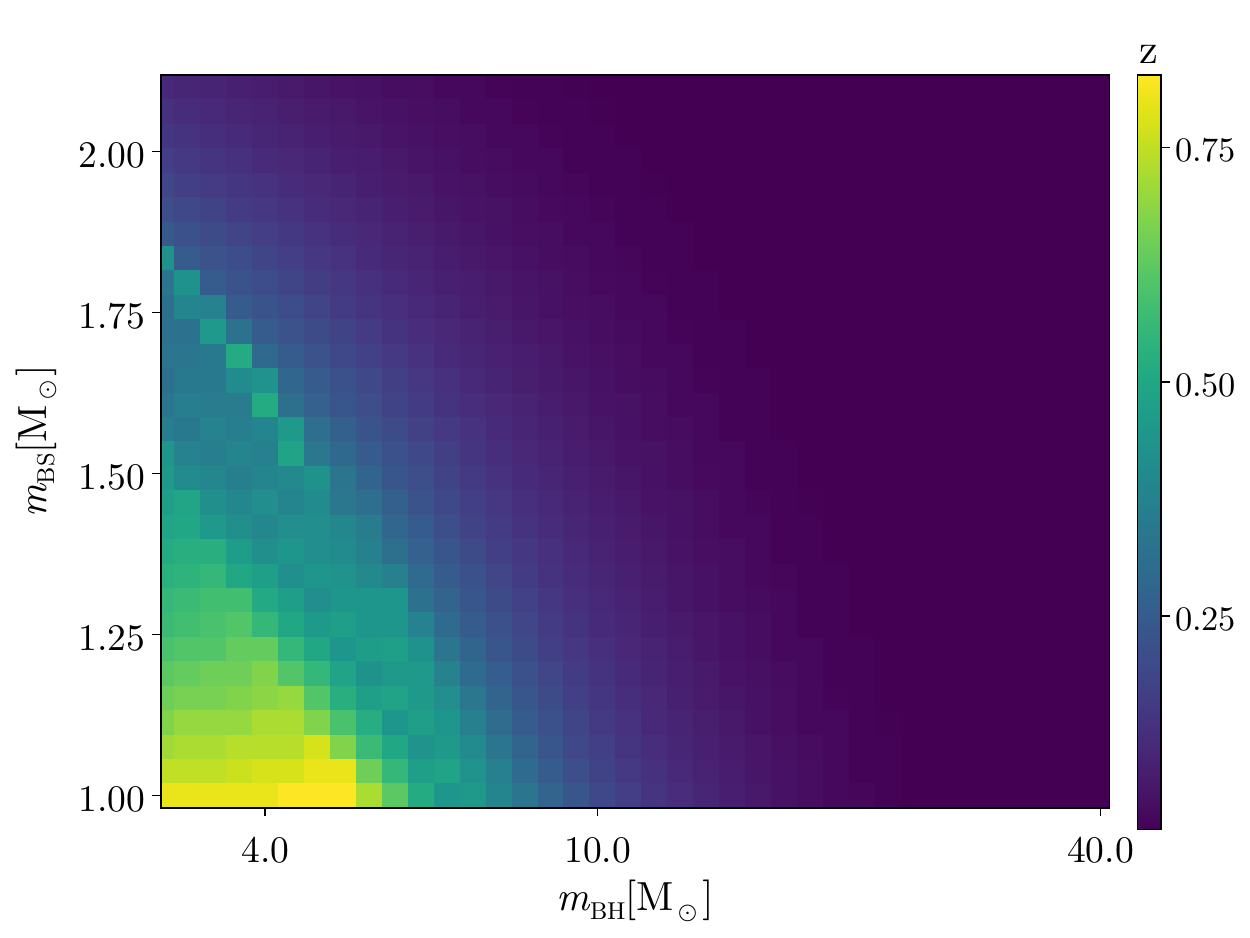}
        \caption{Same plot as before in the case of a BH as the first object. The second object is a NS and the waveform used is \texttt{IMRPhenomNSBH}. We limit ourselves to masses greater than a solar mass. As before, we constrain the maximum redshift at which we can have a 3$\sigma$ exclusion of the tidal deformability $\Lambda$ of a PBH (\textit{left}) or BS (\textit{right}). Also in this case, the best constraints are for light objects of similar masses $\sim(3, 1) M_\odot$ in the PBH case, while for the BS case, it is at $\sim(4, 1)M_\odot$. }
        \label{fig:max_dL_Lambda_BH}
\end{figure}

\vspace{0.5cm}
\noindent We deal now with the distinguishability of the nature of the ECOs, in particular, we analyze two situations: i) what is the horizon inside which we can distinguish a NS from a PBH, ii) what is the horizon inside which we can distinguish a NS from a BS.
Using the FIM formalism, we generate a BNS or NSBH signal, then we extract the error on the tidal deformability $\Lambda_2$ and compute the maximum redshift at which we can reject the PBH hypothesis or the BS hypothesis with a 3$\sigma$ confidence interval. 
Therefore, we examine these two disequations
\begin{align}\label{eq:tidal}
    \Lambda_2^{\rm AP3} - 3 \sigma(\Lambda_2^{\rm})\vert_{ z = z_{3\sigma}} > \Lambda^{\rm PBH}\\
    \Lambda_2^{\rm AP3} + 3 \sigma(\Lambda_2^{\rm})\vert_{ z = z_{3\sigma}} < \Lambda^{\rm BS}\,.
\end{align}
Note that $\Lambda^{\rm PBH}=0$,  and we solve for $z_{3\sigma}$ in the two cases. Moreover, $\Lambda^{\rm PBH}< \Lambda^{\rm AP3}< \Lambda^{\rm BS}$ for all masses considered; thus, inside the calculated horizons, the distinguishability between PBH and BS follows. We do not reject the SQSs hypothesis, i.e., attempt to distinguish between SQSs and NS, since the differences in the tidal deformabilities are very small.
These prospects of using the tidal deformability to distinguish the nature of the compact objects are quantified in \cref{fig:max_dL_Lambda_NS} when the first object is a NS, while \cref{fig:max_dL_Lambda_BH} deals with the BH case. We analyze bodies above one solar mass and this choice was made to fulfill the calibration ranges of the chosen waveforms, i.e., \texttt{IMRPhenomD\_NRTidal\_v2} for the BNS and \texttt{IMRPhenomNSBH} for NSBH.
Concerning the first case, we can picture the situation in which an unknown second body merges with a NS, and we constrain the nature of this second body using its tidal deformability. We show on the $y$ axis the relative mass difference between the first body and the second one.
The procedure we follow is the same as figs. 1--4, we plot the median redshift $z_{3\sigma}$ by solving \cref{eq:tidal} on the tidal deformability $\Lambda_2$ and therefore we compute the maximum redshift at which we can reject the PBH hypothesis (\textit{left}) or the BS hypothesis (\textit{right}) with a 3$\sigma$ confidence interval.

An important assumption is that we condition on the tidal deformability of the first NS $\Lambda_1$. This assumes that we have some outside knowledge of the first body, e.g., astronomical observation, or enough knowledge of the EOS of NS to condition on it. We are projecting these results for 3G detectors, so our current understanding of EOS could change drastically in the meantime.
Concerning the results, the tidal effects are discernible for $z \lesssim 0.4$, with $\sim 3\sigma$ significance in the most favorable part of the parameter space, which corresponds to masses $\sim(1.1, 1.1) M_\odot$ in the PBH case. In the right panel, we reject a BS with $\Lambda = 5 \times \Lambda_{\rm AP3}$, extending the detection range to $z\sim2.5$, for masses $\sim(1.4, 1.3)M_\odot$. This happens because the BS considered are significantly deformable, leading to optimistic constraints using our setup. The picture changes significantly if we remove the conditioning on the first tidal parameter, due to the large correlation of the two tidal deformations. In this case, a better constraint could be obtained on the combination $\tilde\Lambda$.

In \cref{fig:max_dL_Lambda_BH}, we follow the same procedure as before, with the important difference that the first object is a BH thus no conditioning is needed. The waveform used is \texttt{IMRPhenomNSBH} and we look for a $3\sigma$ exclusion of a PBH Hypothesis (\textit{left}) and BS (\textit{right}).
The best constraints are achieved for small mass BHs and small mass NSs, masses $\sim(3, 1) M_\odot$ in the PBH case, while for the BS case, it is at $\sim(4, 1)M_\odot$. 
Thus, this figure illustrates a key limitation: while mass-based PBH identification is viable at cosmological distances, tidal discrimination is restricted to the local Universe, unless some additional information can be incorporated to perform the conditioning procedure followed in \cref{fig:max_dL_Lambda_NS}.

\begin{figure}
    \centering
    \includegraphics[width=\linewidth]{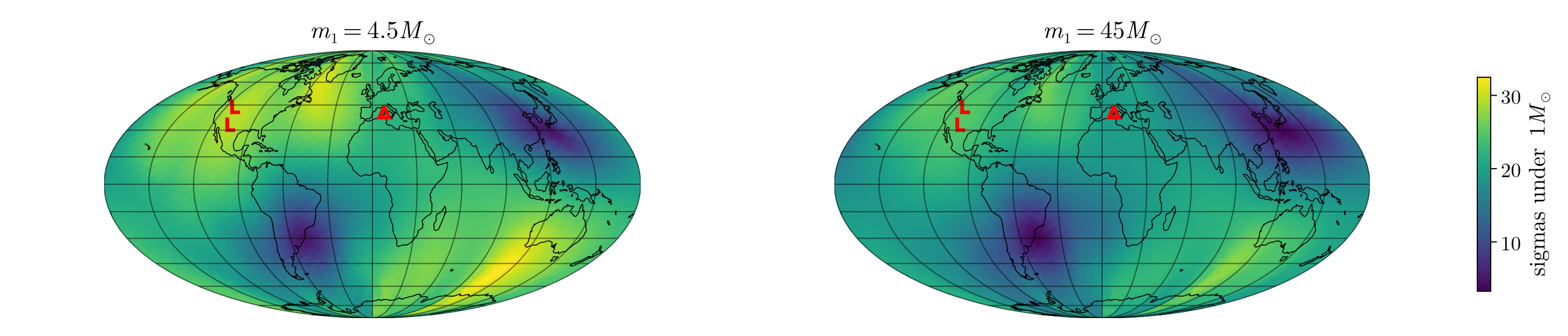}
    \caption{Number of sigmas for a subsolar detection of a PBH. The plots show the detector frame, i.e., a frame in which we observe the sky rotating with a period of one day. The detectors used are represented by red signs, L's for the two CE and $\Delta$ for the triangular ET.
    The event chosen has a PBH mass $m_{\rm PBH} = 0.3 M_\odot$ and a primary mass of $m_{\rm BH} = 4.5 M_\odot$ (\textit{left}) and $m_{\rm BH} = 45 M_\odot$ (\textit{right}). The sky location leads to largely different results spanning from a $\sim 3-4$ sigma measurement to a more than 30 sigma measurement.} 
    \label{fig:skyplot_sigma_subsolar}
\end{figure}
In fact, tidal deformability enters the waveform only at 5PN order and becomes significant primarily in the late inspiral. Even with 3G sensitivity, this restricts tidal measurements to nearby, high-SNR events.  
For BNS systems (\cref{fig:max_dL_Lambda_NS}), confident exclusion of a PBH secondary requires  
$z \lesssim 0.3$--$0.4$ in the most favorable mass ranges.  
For NSBH systems (fig.~6), the reach is similar, with the strongest constraints when both objects 
are on the lighter end of the relevant mass spectrum. Discrimination against a BS companion extends somewhat farther, but still only to $z \lesssim \mathcal{O}(1)$.  
Thus, while mass-based identification is possible to cosmological distances, establishing the internal structure of a light compact object is inevitably confined to the local universe.

So far, we tested the different scenarios this paper aimed to study, evaluating the horizons for a $3\sigma$ detection. We now check the influence of sky position on the detection of a sub-solar object. This test is useful to understand why the procedure of taking the median LD $d_{L,\,3\sigma}$ was important. Here, in fact, we focus on the sky position, but similar cases could be examined for the other extrinsic parameters. We aim to show how the same event can lead to very different detection thresholds when the position on the sky changes, remarking the role of studying catalogs of events to draw meaningful conclusions. In particular, the influence of sky position on sub-solar mass detection is presented in \cref{fig:skyplot_sigma_subsolar}. For a fiducial PBH mass of $0.3\,M_\odot$ and BH companions of $4.5\,M_\odot$ (\textit{left}) and $45\,M_\odot$ (\textit{right}), we observe dramatic variations in detection significance due to the directional sensitivity of the 3G detector network. At best, over $30\sigma$ detection is possible; at worst, significance dips to $\sim 3\sigma$, indicating that the significance of a future detection will be heavily influenced by the sky position. The redshift used for the left panel event is $z=0.65$ and the two events have similar SNR $\sim 10$ after being averaged over the sphere.

\begin{figure}[h!]
    \centering
    \includegraphics[width=0.5\linewidth]{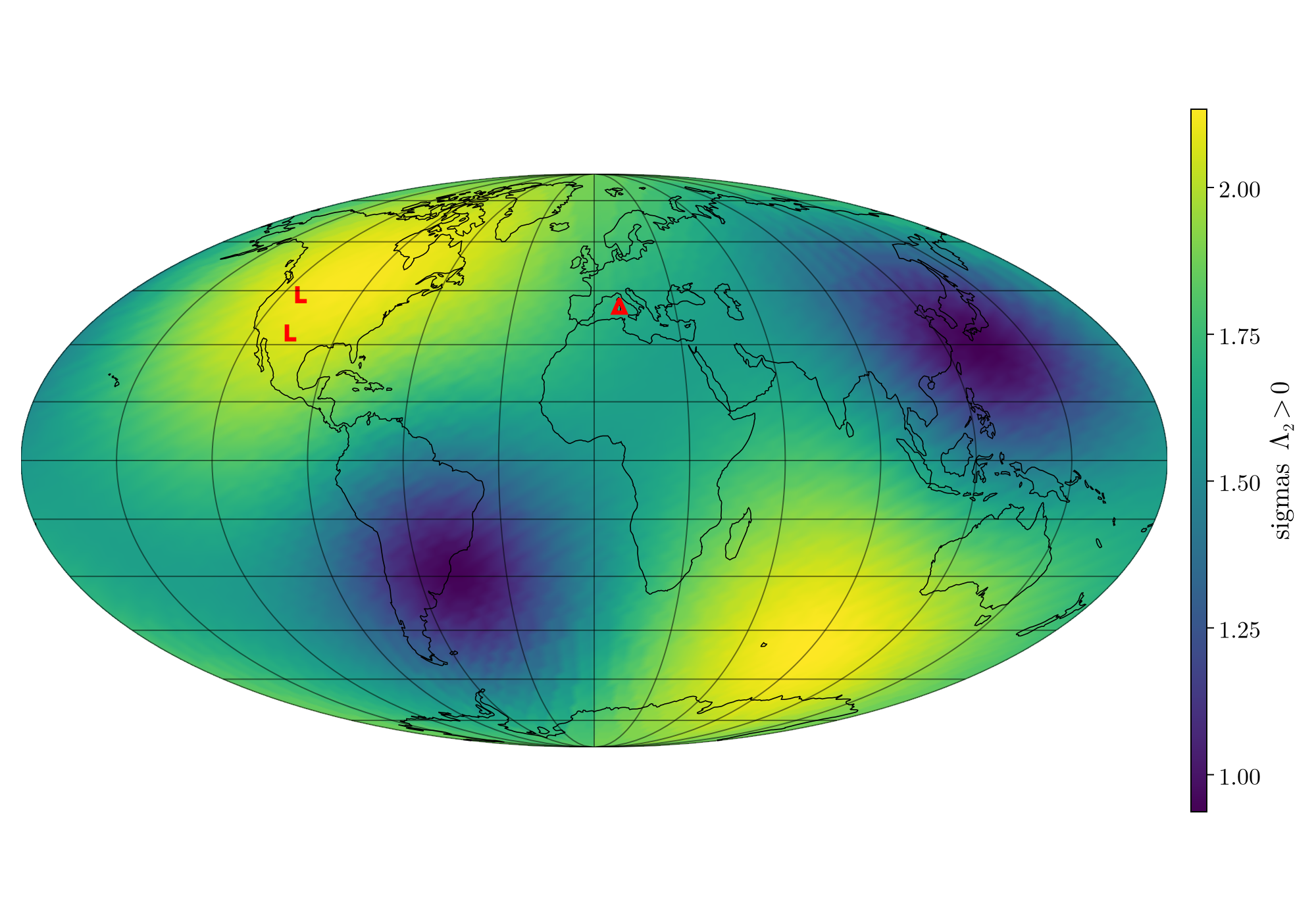}
    \caption{Number of sigmas to claim a $\Lambda_2>0$. The events are BNS obtained with \texttt{IMRPhenomD\_NRTidal\_v2} and we follow the same procedure detailed before, i.e., we condition on the tidal deformability of the first NS $\Lambda_1$. In this case, there is a less significant difference with respect to \cref{fig:skyplot_sigma_subsolar} in the number of sigmas achieved in different parts of the sky, ranging in $\sim [1, 2] \sigma$. The results resemble those of \cref{fig:skyplot_sigma_subsolar} with a significant smearing of its sharper feature.}
    \label{fig:skyplot_sigma_lambda}
\end{figure}

Moreover, we examine the effect of the sky-position for ruling out the PBH hypothesis with tidal effects. We show it in \cref{fig:skyplot_sigma_lambda} and we reproduce the procedure used for \cref{fig:max_dL_Lambda_NS}.
The statistical significance of detecting $\Lambda_2 > 0$ hovers between $1\sigma$ and $2\sigma$, depending on the sky position. The angular modulation resembles that seen in sub-solar mass detection, but with lower overall significance.

Figs.~7--8 show that sky location produces large modulations in the measured significance of both 
mass-based identification and tidal discrimination.  
For sub-solar identification (fig.~7), the significance can range from $\sim 3\sigma$ to 
$\gtrsim 30\sigma$ for the same intrinsic binary at fixed redshift, depending solely on sky position.  
Tidal effects (fig.~8), being weaker and more localized to the late inspiral, show smaller variation 
but remain sensitive to directional differences.  
These modulations imply that the probability of achieving strong discrimination for a given binary 
depends not only on its masses and redshift, but also on its location relative to the network’s 
antenna response.

\begin{figure}[h!]
    \centering
    \includegraphics[width=0.5\linewidth]{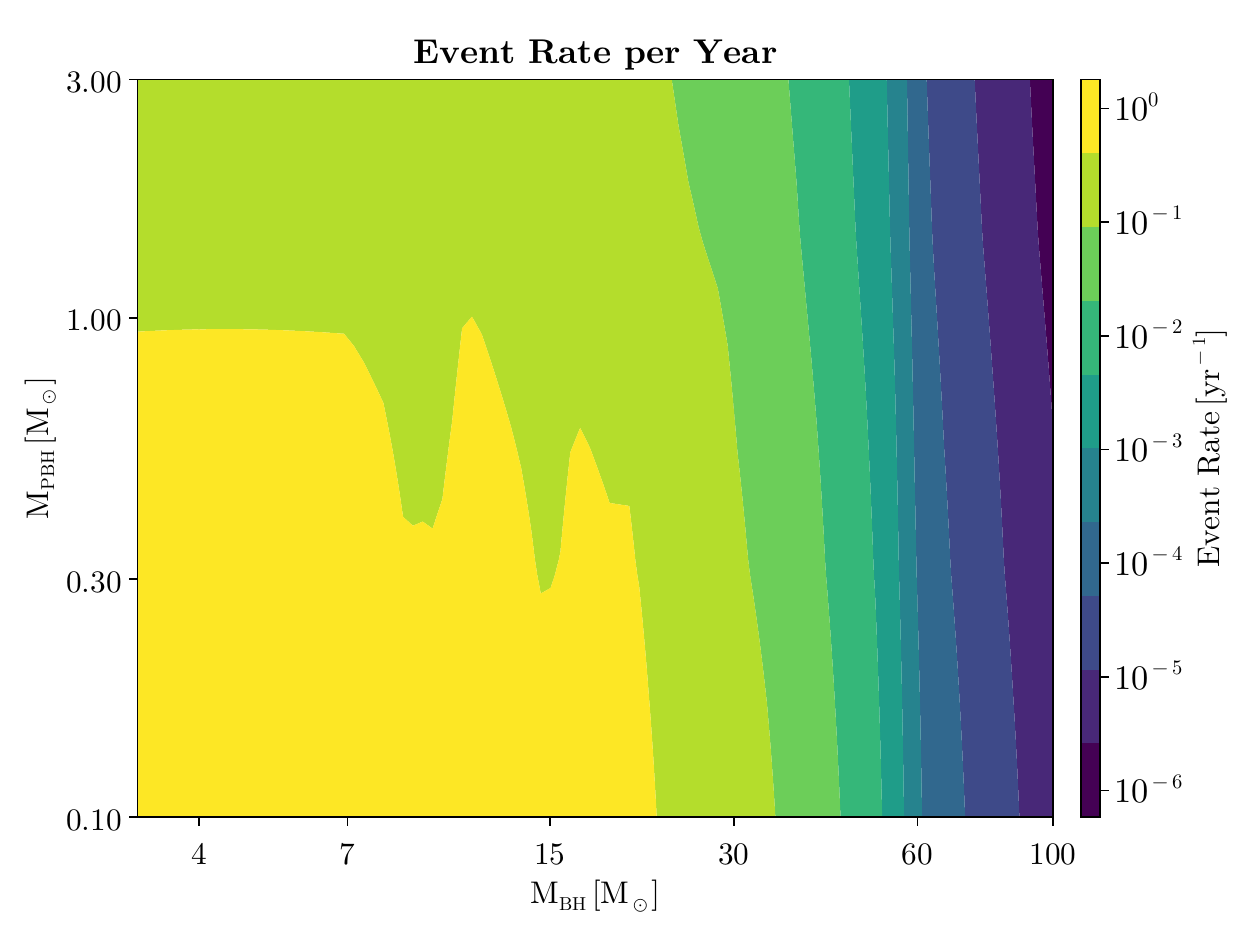}
    \includegraphics[width=0.49\linewidth]{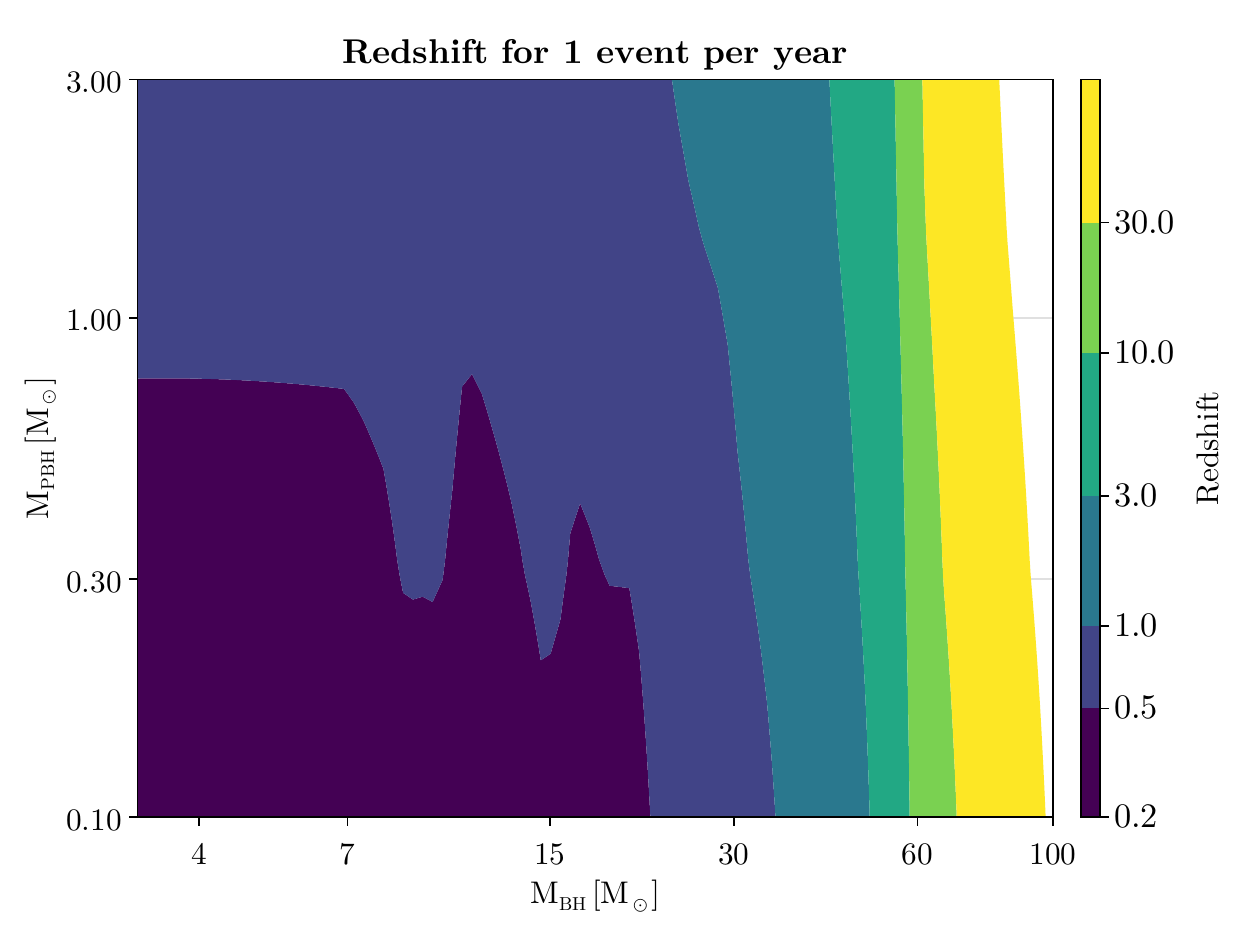}
    \caption{Event rate within a one Gpc radius sphere (\textit{left}) and redshift corresponding to one event per year (\textit{right}) of stellar BH vs PBH. Lighter mergers are preferred and, importantly, most of the interesting parameter space from \cref{fig:max_dL_mass_subsolar_BH} has at least one event a year.}
    \label{fig:eventrate}
\end{figure}

Finally, \cref{fig:eventrate} summarizes PBH detection prospects in terms of expected event rates and redshift reach. We solely consider mergers of stellar black holes (whose mass is on the $x$ axis) with PBH ($y$ axis) in the mass range between 0.1 and 3 solar masses. For the former, we adopt the stellar black hole mass function as modeled by Sicilia et al.~\cite{Sicilia:2021gtu}, which computes an {\it ab initio} relic distribution using the SEVN stellar/binary evolution code alongside galaxy formation prescriptions; for the PBH abundance, we instead assume, for a given PBH mass, the maximal-possible abundance of PBHs, i.e., we maximize $f_{\rm PBH}$, assuming a monochromatic mass function, and using the constraints of Ref.~\cite{Poulin:2017pbh, Serpico:2020pbh}. The merger rate functional form we employ follows an analytic estimate of the fitting expression provided in
Ref.~\cite{Stasenko:2024pbh}, suitable for phenomenological studies. The dips visible between PBH masses of 0.3 and 1 solar masses are due to the shape of the constraints on the PBH abundance $f_{\rm PBH}(M_{\rm PBH})$ in that mass range.

The left panel shows contours of 1 event per year within a sphere of 1 Gpc radius on the plane of stellar-mass PBH vs light PBH. The right panel, instead, shows the distance, expressed as a redshift, within which one event per year is expected: in other words, for a given choice of the PBH and astrophysical black hole mass, the contours indicate the redshift within which one event per year or more is expected. 
These figures define the observational frontier for probing the nature of light exotic compact objects using third-generation gravitational-wave observatories.

Note that since 1 Gpc approximately corresponds to $z\simeq 0.23-0.24$, \cref{fig:eventrate} indicates that in a large swath of parameter space there may be as many as ${\cal O}(10)$ events involving a sub-solar mass object where the mass can be confidently said to be sub-solar and will be detectable with 3G detectors.  The left panel shows that mergers in the most favorable mass range 
(e.g., $30\,M_\odot$--$0.3\,M_\odot$) occur at rates of $\mathcal{O}(1)$ per year within a gigaparsec.

\section{Discussion and Conclusions}
\label{sec:conclusions}

In this work, we have investigated the prospects for identifying sub-solar mass primordial black holes and distinguishing them from possible astrophysical and exotic compact object impostors using third-generation gravitational-wave detectors such as the Einstein Telescope and Cosmic Explorer. Employing the Fisher matrix formalism, we quantified the maximum redshifts at which sub-solar mass companions can be measured with high significance, as well as the redshift ranges where tidal deformability measurements could exclude the primordial black hole hypothesis.

Our analysis shows that sub-solar mass primordial black holes, if present in binary systems with neutron stars or stellar-mass black holes, will be detectable out to cosmological distances. In particular, we find that black hole–primordial black hole binaries can be observed up to redshift $z \gtrsim 3$, and in some configurations even farther, while neutron star–primordial black hole binaries are typically detectable out to redshift $z \sim 1$. These results establish that third-generation detectors will have the sensitivity required to probe sub-solar primordial black holes well beyond the capabilities of the current LIGO–Virgo–KAGRA network.

A central challenge, however, lies in establishing the nature of the detected low-mass compact objects. While primordial black holes are characterized by vanishing tidal deformability ($k_2 = 0$), neutron stars, strange quark stars, and boson stars exhibit finite tidal signatures that depend on their respective equations of state. We have shown that tidal measurements, though in principle decisive, are restricted to the local Universe: even with third-generation sensitivity, confident exclusion of the primordial black hole hypothesis through detection of non-zero tidal effects is possible only for redshifts $z \lesssim 0.3$--$0.5$, depending on the equation of state and binary configuration. This implies that while mass-based identification of sub-solar objects as primordial black holes can be achieved at high redshift, tidal-based discrimination between primordial black holes and material impostors is fundamentally limited to nearby events.

Our results highlight a dual observational frontier. On the one hand, cosmological reach for sub-solar mass detections offers unprecedented opportunities to map the primordial black hole parameter space, probe early-Universe physics, and test the hypothesis that primordial black holes constitute a fraction of the dark matter. On the other hand, the limited redshift horizon for tidal discrimination underscores the importance of low-redshift, high signal-to-noise events for definitively ruling out non-primordial interpretations. This complementarity suggests that robust population-level inference strategies--combining mass spectra, spin distributions, event rates, and tidal signatures--will be essential for establishing the primordial origin of light black holes.

The analysis presented here relies on several simplifying assumptions, including the use of phenomenological prescriptions for strange quark star and boson star equations of state, waveform models calibrated within limited mass-ratio regimes, and the Fisher matrix approximation. Future work should refine these aspects, for instance by employing Bayesian parameter estimation with full inspiral–merger–ringdown waveforms tailored to exotic compact objects, by incorporating more realistic population models of primordial black holes and exotic compact objects, and by exploring synergies with electromagnetic and cosmological probes.

In conclusion, we find that third-generation gravitational-wave observatories will provide a decisive test on the existence of sub-solar primordial black holes. The detection of a sub-solar mass black hole would be a striking signal of new physics, strongly indicative of a primordial origin. Conversely, detection of finite tidal effects in the same mass range would point toward the realization of exotic states of matter, such as strange quark matter or bosonic condensates. Either outcome would have profound implications for nuclear physics, particle physics, and cosmology, reinforcing the role of gravitational waves as a unique tool for exploring the deep connections between astrophysics and fundamental physics.

\appendix
\section{Technical details}\label{app}

\subsection{Fisher formalism}\label{sec:formalism}
The output of an interferometer is a datastream $d(t)$, which can be expressed as  
\begin{equation}\label{eq:signal}
    d(t) = s(t) + n(t)\,,
\end{equation}
where $s(t)$ represents the GW signal and $n(t)$ represents the noise, which is characterized by the detector PSD $S_n(f)$. In fact, after going into Fourier space, one can write
\begin{equation}
    \langle n(f)\,n^*(f')\rangle=\frac{1}{2}\delta(f-f')S_n(f)\,.
\end{equation}
where $n(f)$ represents the noise in the frequency domain and the symbol $*$ is the complex conjugate.
We also introduce the scalar-product $(a\mid b)$ \cite{maggiore2008gravitational}
\begin{equation}\label{eq:scalar_prod}
    (a \mid b)=2 \int_{f_{\rm min}}^{f_{\rm max}} \frac{a(f) b^*(f)+a^*(f) b(f)}{S_n(f)} \mathrm{d} f,
\end{equation}
which is a noise-weighted scalar product between the minimum frequency $f_{\rm min}$ the maximum frequency $f_{\rm max}$ which depend on the detector PSD and the maximum frequency of the signal. 
Before introducing the FIM, we introduce the Signal-to-Noise, which gives a measure of the strength of the signal and, for a single detector, is defined as
    \begin{equation}
        {\rm SNR} = (s\mid s)^{1/2}\,,
    \end{equation}
while for a network of detectors, the SNR is
    \begin{equation}\label{eq:network_SNR}
        {\rm SNR}_{network} = \left[\sum_i {\rm SNR}_i^2\right]^{1/2}\,,
    \end{equation}
    where $i$ indicates the different detectors. In this work, we consider a signal to be detected when the network SNR is above a threshold, which we take to be $\rm SNR_{thres}$ = 8.\\
    Under the assumption that the noise is stationary, Gaussian distributed and with zero mean, the FIM defined as in Eq. \eqref{eq:FIM}, which in this notation becomes
      \begin{equation}
       \Gamma_{i j}=\left(\left.\frac{\partial s}{\partial \theta^i} \right\rvert\, \frac{\partial s}{\partial \theta^j}\right)\,,
    \end{equation}
    where $\theta^i$ represents one of the CBC parameters $\boldsymbol{\theta}$ and $\log\mathcal{L}$ is the standard Gaussian likelihood (see e.g.~\cite{Rodriguez:2013mla}) of the datastream $d$ measured at the detector given the event parameters $\boldsymbol{\theta}$. Moreover, $\langle\dots\rangle_n$ indicates the ensemble average over the noise realizations.

\subsection{Detectors}
The positions and orientations of the detectors are reported in \cref{tab:positions} while their power spectral densities are plotted in the \textit{left} panel of \cref{fig:app_plot}.
\begin{table}[h]    
    \centering
\begin{tabular}{|c|c|c|c|}
\hline
 & Latitude & Longitude & Orientation  \\
\hline
CE 40 km & 43.83 & -112.82 & -45.0 \\
\hline
CE 20 km & 33.16 & -106.48 & -105.0 \\
\hline
ET 10 km triangular & 40.52 & 9.42 & 0.0 \\
\hline
\end{tabular}
\caption{Latitude, longitude and orientation with respect to the local East of the three detectors used in this work.}
\label{tab:positions}
\end{table}

\subsection{Tidal deformabilities}
\begin{figure}
    \centering
    \includegraphics[width=0.5\linewidth]{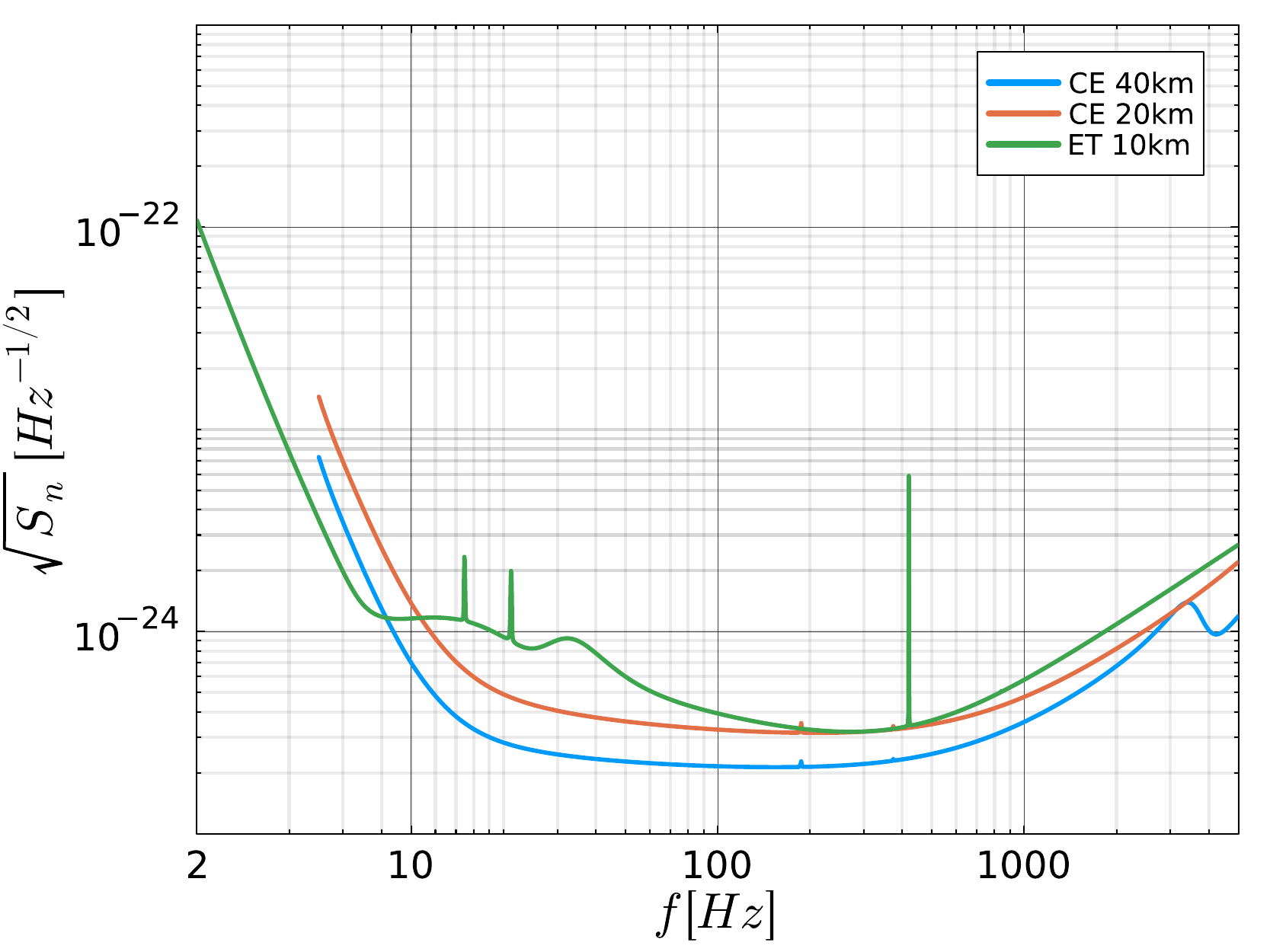}
    \includegraphics[width=0.49\linewidth]{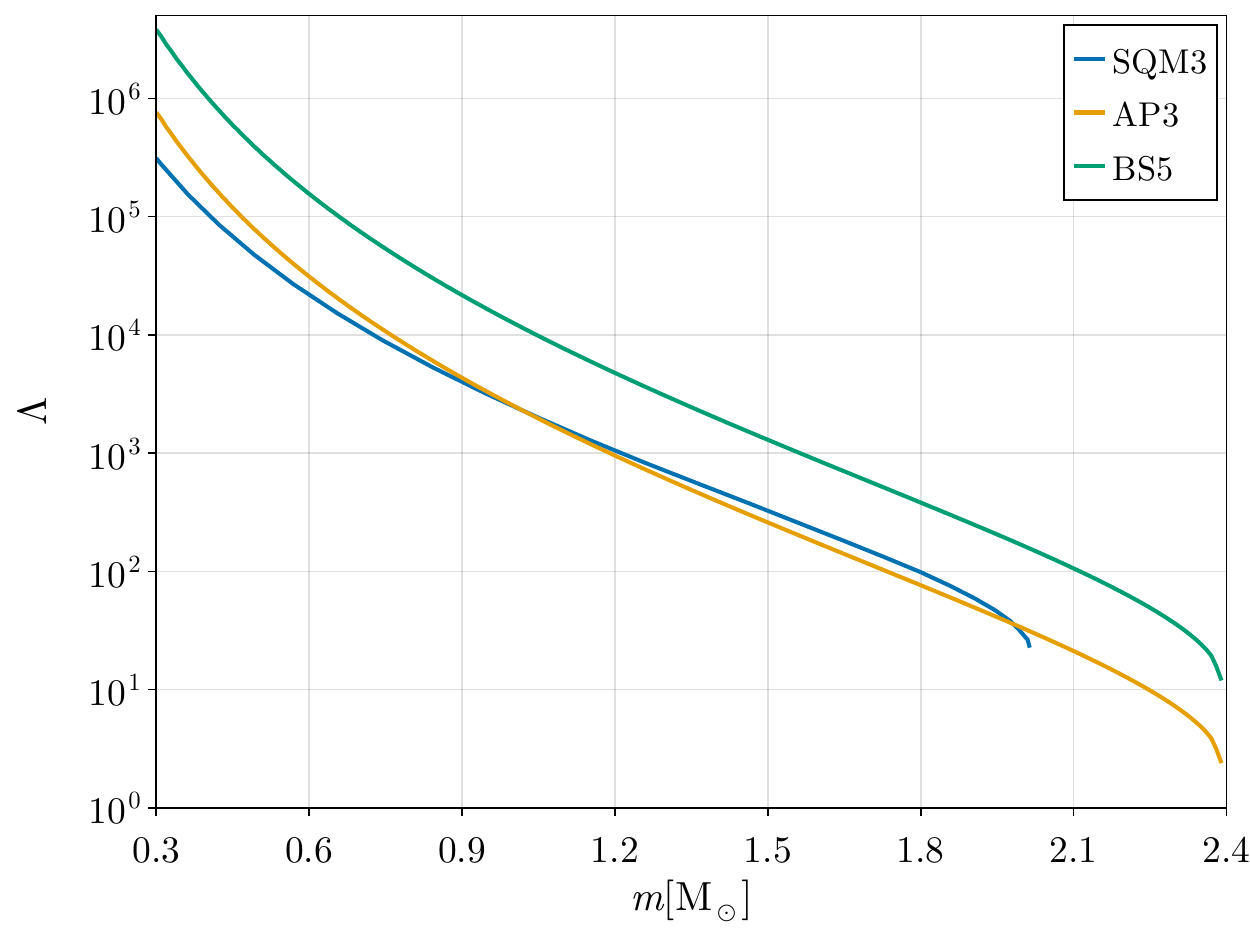}
    \caption{\textit{Left}: PSD of the three detectors used in this work, one L-shape Cosmic Explorer with 40 km arm lengths (blue), one L-shape Cosmic Explorer with 20 km arm lengths (orange) and a triangular Einstein Telescope with 10km arms (green)\\
    \textit{Right}: Relations of mass of the compact object $m$ and their tidal deformability $\Lambda$ for a strange quark star SQM3 (blue), for a neutron star with EOS AP3 (orange) and for the phenomenological boson star model employed in this work BS5 (green).}
    \label{fig:app_plot}
\end{figure}

The mass of the bosonic constituent $m_b$ sets the fundamental mass and length scales of a boson star, 
with the maximum mass scaling as $M_{\max} \sim M_{\rm Pl}^2/m_b$ and the typical radius $R \sim (G m_b^2 M)^{-1}$~\cite{Kaup:1968zz,Ruffini:1969qy}. 
The effective EOS arises from the scalar potential: for free bosons the pressure--density relation is entirely determined by the field configuration, 
while with quartic self-interactions, $V(\phi) = \frac{\lambda}{4}|\phi|^4$, one can approximate a polytropic form $p \sim (\lambda/4m_b^4)\rho^2$~\cite{Colpi:1986ye}.
Given an assumed EOS, the mass--radius relation $M(R)$ is fixed, and thus the compactness $C=GM/R$, which controls the dimensionless tidal deformability 
$\Lambda = \tfrac{2}{3}k_2 C^{-5}$, with $k_2$ the Love number~\cite{Hinderer:2007mb}. 
Lighter bosons generically lead to extended, low-compactness configurations with large $\Lambda$, 
while heavier bosons or strongly self-interacting fields yield more compact stars with suppressed tidal signatures, 
potentially compatible with gravitational-wave constraints from events such as GW170817~\cite{Abbott:2018exr}. In our study, we do not impose any theoretical priors on $\lambda$ and $m_b$ and the resulting microphysics, but rather assume a phenomenological EOS. In \cref{fig:app_plot}, we show the three EOS used in this work. The SQM3 tidal deformability~\cite{wang2021tidal} is similar to the AP3 one, with the significant difference that SQM3 has a maximum mass $M_{\rm max} \sim 2\,M_\odot$, while for AP3 $M_{\rm max} \sim 2.4\,M_\odot$. As said in the main text, the BS5 curve reproduces five times the deformability of a AP3 tidal deformability.

\begin{acknowledgments}
This work is partly supported by the U.S.\ Department of Energy grant number de-sc0010107 (SP). 
AB is supported by ICSC – Centro Nazionale di Ricerca in High Performance Computing, Big Data and Quantum Computing, funded by European Union – NextGenerationEU”. AB would like to thank Stefano Anselmi, Mauro Pieroni, Alessandro Renzi and Angelo Ricciardone for early discussions on the project. AB would like to thank Juan Garcia-Bellido and Sachsa Husa for useful discussions.
The authors would also like to thank Liam Colombo Murphy for the strange quark stars EOS solver. The authors would like to thank the
anonymous referee for helpful suggestions.
\end{acknowledgments}


\bibliographystyle{JHEP}

\bibliography{biblio.bib}

@article{Postnikov:2010yn,
  author = {Postnikov, S. and Prakash, M. and Lattimer, J. M.},
  title = {Tidal Love numbers of neutron and self-bound quark stars},
  journal = {Phys. Rev. D},
  volume = {82},
  pages = {024016},
  year = {2010},
  doi = {10.1103/PhysRevD.82.024016}
}

@article{Begnoni:2025oyd,
    author = "Begnoni, Andrea and Anselmi, Stefano and Pieroni, Mauro and Renzi, Alessandro and Ricciardone, Angelo",
    title = "{Detectability and Parameter Estimation for Einstein Telescope Configurations with GWJulia}",
    eprint = "2506.21530",
    archivePrefix = "arXiv",
    primaryClass = "astro-ph.CO",
    month = "6",
    year = "2025"
}

@article{DeLuca:2024,
    author = "De Luca, Valerio and Franciolini, Gabriele and Riotto, Antonio",
    title = "{Flea on the elephant: Tidal Love numbers in subsolar primordial black hole searches}",
    eprint = "2408.14207",
    archivePrefix = "arXiv",
    primaryClass = "gr-qc",
    reportNumber = "CERN-TH-2024-140",
    doi = "10.1103/PhysRevD.110.104041",
    journal = "Phys. Rev. D",
    volume = "110",
    number = "10",
    pages = "104041",
    year = "2024"
}

@article{Zhou:2017pha,
    author = "Zhou, En-Ping and Zhou, Xia and Li, Ang",
    title = "{Constraints on interquark interaction parameters with GW170817 in a binary strange star scenario}",
    eprint = "1711.04312",
    archivePrefix = "arXiv",
    primaryClass = "astro-ph.HE",
    doi = "10.1103/PhysRevD.97.083015",
    journal = "Phys. Rev. D",
    volume = "97",
    number = "8",
    pages = "083015",
    year = "2018"
}

@article{Zhiqiang:2021,
    author = "Miao, Zhiqiang and Jiang, Jin-Liang and Li, Ang and Chen, Lie-Wen",
    title = "{Bayesian Inference of Strange Star Equation of State Using the GW170817 and GW190425 Data}",
    eprint = "2107.13997",
    archivePrefix = "arXiv",
    primaryClass = "astro-ph.HE",
    doi = "10.3847/2041-8213/ac194d",
    journal = "Astrophys. J. Lett.",
    volume = "917",
    number = "2",
    pages = "L22",
    year = "2021"
}

@article{Vines:2011cz,
  author = {Vines, J. and Flanagan, {\'E}.E.},
  title = {Post-1-Newtonian tidal effects in the gravitational waveform from binary inspirals},
  journal = {Phys. Rev. D},
  volume = {83},
  pages = {084051},
  year = {2011},
  doi = {10.1103/PhysRevD.83.084051}
}

@article{Dupletsa:2022scg,
    author = "Dupletsa, Ulyana and Harms, Jan and Banerjee, Biswajit and Branchesi, Marica and Goncharov, Boris and Maselli, Andrea and Oliveira, Ana Carolina Silva and Ronchini, Samuele and Tissino, Jacopo",
    title = "{gwfish: A simulation software to evaluate parameter-estimation capabilities of gravitational-wave detector networks}",
    eprint = "2205.02499",
    archivePrefix = "arXiv",
    primaryClass = "gr-qc",
    doi = "10.1016/j.ascom.2022.100671",
    journal = "Astron. Comput.",
    volume = "42",
    pages = "100671",
    year = "2023"
}

@article{Perot:2023ghi,
    author = {Perot, Lo{\"\i}c and Chamel, Nicolas},
    title = "{Role of Quark Matter and Color Superconductivity in the Structure and Tidal Deformability of Strange Dwarfs}",
    doi = "10.3390/universe9090382",
    journal = "Universe",
    volume = "9",
    number = "9",
    pages = "382",
    year = "2023"
}

@article{Bhattacharyya:2016nhb,
    author = "Bhattacharyya, Sudip and Bombaci, Ignazio and Logoteta, Domenico and Thampan, Arun V.",
    title = "{Fast spinning strange stars: possible ways to constrain interacting quark matter parameters}",
    eprint = "1601.06120",
    archivePrefix = "arXiv",
    primaryClass = "astro-ph.HE",
    doi = "10.1093/mnras/stw206",
    journal = "Mon. Not. Roy. Astron. Soc.",
    volume = "457",
    number = "3",
    pages = "3101--3114",
    year = "2016"
}

@article{Witten:1984rs,
    author = "Witten, Edward",
    title = "{Cosmic Separation of Phases}",
    journal = "Phys. Rev. D",
    volume = "30",
    pages = "272-285",
    year = "1984",
    doi = "10.1103/PhysRevD.30.272"
}

@article{Zdunik:2000xx,
    author = "Zdunik, J. L.",
    title = "{On the mass of moderately rotating strange stars}",
    journal = "Astron. Astrophys.",
    volume = "359",
    pages = "311-314",
    year = "2000",
    eprint = "astro-ph/0004375"
}

@article{Yoshida:1997qf,
    author = "Yoshida, Shijun and Eriguchi, Yoshifumi",
    title = "{Rotating boson stars in general relativity}",
    journal = "Phys. Rev. D",
    volume = "56",
    year = "1997",
    pages = "762-771",
    doi = "10.1103/PhysRevD.56.762",
    archivePrefix = "arXiv",
    primaryClass = "gr-qc"
}

@article{Colpi:1986ye,
    author = "Colpi, Monica and Shapiro, Stuart L. and Wasserman, Ira",
    title = "{Boson Stars: Gravitational Equilibria of Selfinteracting Scalar Fields}",
    journal = "Phys. Rev. Lett.",
    volume = "57",
    year = "1986",
    pages = "2485-2488",
    doi = "10.1103/PhysRevLett.57.2485"
}

@article{ET:2019Magg,
    author = "Maggiore, Michele and others",
    collaboration = "ET",
    title = "{Science Case for the Einstein Telescope}",
    eprint = "1912.02622",
    archivePrefix = "arXiv",
    primaryClass = "astro-ph.CO",
    doi = "10.1088/1475-7516/2020/03/050",
    journal = "JCAP",
    volume = "03",
    pages = "050",
    year = "2020"
}

@article{Lattimer2004,
    author = "Lattimer, J. M. and Prakash, M.",
    title = "{The physics of neutron stars}",
    eprint = "astro-ph/0405262",
    archivePrefix = "arXiv",
    doi = "10.1126/science.1090720",
    journal = "Science",
    volume = "304",
    pages = "536--542",
    year = "2004"
}

@misc{lalsuite,
       author         = "{LIGO Scientific Collaboration} and {Virgo Collaboration} and {KAGRA Collaboration}",
       title          = "{LVK} {A}lgorithm {L}ibrary - {LALS}uite",
       howpublished   = "Free software (GPL)",
       doi            = "10.7935/GT1W-FZ16",
       year           = "2018"
 }

@article{Benvenuto1998,
    author = "Benvenuto, O. G. and Lugones, G.",
    title = "{The properties of strange stars in the quark mass-density-dependent model}",
    doi = "10.1142/S0218271898000048",
    journal = "Int. J. Mod. Phys. D",
    volume = "7",
    pages = "29--48",
    year = "1998"
}

@article{Cromartie:2019kug,
    author = "Cromartie, H. T. and et al.",
    title = "{Relativistic Shapiro delay measurements of an extremely massive millisecond pulsar}",
    journal = "Nature Astronomy",
    volume = "4",
    pages = "72-76",
    year = "2020",
    doi = "10.1038/s41550-019-0880-2",
    eprint = "1904.06759",
    archivePrefix = "arXiv",
    primaryClass = "astro-ph.HE"
}

@article{Schunck:1999zu,
    author = "Schunck, Franz E. and Mielke, Eckehard W.",
    title = "{General relativistic boson stars}",
    journal = "Class. Quant. Grav.",
    volume = "20",
    pages = "R301-R356",
    year = "2003",
    doi = "10.1088/0264-9381/20/20/201",
    eprint = "0801.0307",
    archivePrefix = "arXiv",
    primaryClass = "astro-ph"
}

@article{Garcia-Quiros:2020XHM,
    author = "Garc\'\i{}a-Quir\'os, Cecilio and Colleoni, Marta and Husa, Sascha and Estell\'es, H\'ector and Pratten, Geraint and Ramos-Buades, Antoni and Mateu-Lucena, Maite and Jaume, Rafel",
    title = "{Multimode frequency-domain model for the gravitational wave signal from nonprecessing black-hole binaries}",
    eprint = "2001.10914",
    archivePrefix = "arXiv",
    primaryClass = "gr-qc",
    doi = "10.1103/PhysRevD.102.064002",
    journal = "Phys. Rev. D",
    volume = "102",
    number = "6",
    pages = "064002",
    year = "2020"
}

@article{Husa:2015PhD,
    author = {Husa, Sascha and Khan, Sebastian and Hannam, Mark and P\"urrer, Michael and Ohme, Frank and Jim\'enez Forteza, Xisco and Boh\'e, Alejandro},
    title = "{Frequency-domain gravitational waves from nonprecessing black-hole binaries. I. New numerical waveforms and anatomy of the signal}",
    eprint = "1508.07250",
    archivePrefix = "arXiv",
    primaryClass = "gr-qc",
    doi = "10.1103/PhysRevD.93.044006",
    journal = "Phys. Rev. D",
    volume = "93",
    number = "4",
    pages = "044006",
    year = "2016"
}

@article{Khan:2015PhD,
    author = {Khan, Sebastian and Husa, Sascha and Hannam, Mark and Ohme, Frank and P\"urrer, Michael and Jim\'enez Forteza, Xisco and Boh\'e, Alejandro},
    title = "{Frequency-domain gravitational waves from nonprecessing black-hole binaries. II. A phenomenological model for the advanced detector era}",
    eprint = "1508.07253",
    archivePrefix = "arXiv",
    primaryClass = "gr-qc",
    doi = "10.1103/PhysRevD.93.044007",
    journal = "Phys. Rev. D",
    volume = "93",
    number = "4",
    pages = "044007",
    year = "2016"
}

@article{Dietrich:2019NRTidal,
    author = "Dietrich, Tim and Samajdar, Anuradha and Khan, Sebastian and Johnson-McDaniel, Nathan K. and Dudi, Reetika and Tichy, Wolfgang",
    title = "{Improving the NRTidal model for binary neutron star systems}",
    eprint = "1905.06011",
    archivePrefix = "arXiv",
    primaryClass = "gr-qc",
    doi = "10.1103/PhysRevD.100.044003",
    journal = "Phys. Rev. D",
    volume = "100",
    number = "4",
    pages = "044003",
    year = "2019"
}

@article{Pratten:2020XAS,
    author = "Pratten, Geraint and Husa, Sascha and Garcia-Quiros, Cecilio and Colleoni, Marta and Ramos-Buades, Antoni and Estelles, Hector and Jaume, Rafel",
    title = "{Setting the cornerstone for a family of models for gravitational waves from compact binaries: The dominant harmonic for nonprecessing quasicircular black holes}",
    eprint = "2001.11412",
    archivePrefix = "arXiv",
    primaryClass = "gr-qc",
    reportNumber = "LIGO-P2000018",
    doi = "10.1103/PhysRevD.102.064001",
    journal = "Phys. Rev. D",
    volume = "102",
    number = "6",
    pages = "064001",
    year = "2020"
}

@article{Pannarale:2015NSBH,
    author = "Pannarale, Francesco and Berti, Emanuele and Kyutoku, Koutarou and Lackey, Benjamin D. and Shibata, Masaru",
    title = "{Aligned spin neutron star-black hole mergers: a gravitational waveform amplitude model}",
    eprint = "1509.00512",
    archivePrefix = "arXiv",
    primaryClass = "gr-qc",
    reportNumber = "LIGO-P1500135",
    doi = "10.1103/PhysRevD.92.084050",
    journal = "Phys. Rev. D",
    volume = "92",
    number = "8",
    pages = "084050",
    year = "2015"
}

@article{Thompson:2020PhenomNSBH,
    author = "Thompson, Jonathan E. and Fauchon-Jones, Edward and Khan, Sebastian and Nitoglia, Elisa and Pannarale, Francesco and Dietrich, Tim and Hannam, Mark",
    title = "{Modeling the gravitational wave signature of neutron star black hole coalescences}",
    eprint = "2002.08383",
    archivePrefix = "arXiv",
    primaryClass = "gr-qc",
    reportNumber = "LIGO-P2000059",
    doi = "10.1103/PhysRevD.101.124059",
    journal = "Phys. Rev. D",
    volume = "101",
    number = "12",
    pages = "124059",
    year = "2020"
}

@article{Buonanno:2009F2,
    author = "Buonanno, Alessandra and Iyer, Bala and Ochsner, Evan and Pan, Yi and Sathyaprakash, B. S.",
    title = "{Comparison of post-Newtonian templates for compact binary inspiral signals in gravitational-wave detectors}",
    eprint = "0907.0700",
    archivePrefix = "arXiv",
    primaryClass = "gr-qc",
    doi = "10.1103/PhysRevD.80.084043",
    journal = "Phys. Rev. D",
    volume = "80",
    pages = "084043",
    year = "2009"
}

@article{Pan:2007F2,
    author = "Pan, Yi and Buonanno, Alessandra and Baker, John G. and Centrella, Joan and Kelly, Bernard J. and McWilliams, Sean T. and Pretorius, Frans and van Meter, James R.",
    title = "{A Data-analysis driven comparison of analytic and numerical coalescing binary waveforms: Nonspinning case}",
    eprint = "0704.1964",
    archivePrefix = "arXiv",
    primaryClass = "gr-qc",
    doi = "10.1103/PhysRevD.77.024014",
    journal = "Phys. Rev. D",
    volume = "77",
    pages = "024014",
    year = "2008"
}

@article{Boyle:2009F2,
    author = "Boyle, Michael and Brown, Duncan A. and Pekowsky, Larne",
    editor = "Sutton, Patrick and Shoemaker, Deirdre",
    title = "{Comparison of high-accuracy numerical simulations of black-hole binaries with stationary phase post-Newtonian template waveforms for Initial and Advanced LIGO}",
    eprint = "0901.1628",
    archivePrefix = "arXiv",
    primaryClass = "gr-qc",
    doi = "10.1088/0264-9381/26/11/114006",
    journal = "Class. Quant. Grav.",
    volume = "26",
    pages = "114006",
    year = "2009"
}

@article{Gadre:2022sed,
    author = {Gadre, Bhooshan and P{\"u}rrer, Michael and Field, Scott E. and Ossokine, Serguei and Varma, Vijay},
    title = "{Fully precessing higher-mode surrogate model of effective-one-body waveforms}",
    eprint = "2203.00381",
    archivePrefix = "arXiv",
    primaryClass = "gr-qc",
    reportNumber = "LIGO-P2200040",
    doi = "10.1103/PhysRevD.110.124038",
    journal = "Phys. Rev. D",
    volume = "110",
    number = "12",
    pages = "124038",
    year = "2024"
}

@article{Ohme:2011zm,
    author = "Ohme, Frank and Hannam, Mark and Husa, Sascha",
    title = "{Reliability of complete gravitational waveform models for compact binary coalescences}",
    eprint = "1107.0996",
    archivePrefix = "arXiv",
    primaryClass = "gr-qc",
    reportNumber = "LIGO-P1100078-AEI-2011-039",
    doi = "10.1103/PhysRevD.84.064029",
    journal = "Phys. Rev. D",
    volume = "84",
    pages = "064029",
    year = "2011"
}

@article{Lam:2023oga,
    author = "Lam, Kelvin K. H. and Wong, Kaze W. K. and Edwards, Thomas D. P.",
    title = "{Recalibrating a gravitational wave phenomenological waveform model}",
    eprint = "2306.17245",
    archivePrefix = "arXiv",
    primaryClass = "gr-qc",
    doi = "10.1103/PhysRevD.109.124009",
    journal = "Phys. Rev. D",
    volume = "109",
    number = "12",
    pages = "124009",
    year = "2024"
}

@article{Rink:2024swg,
    author = "Rink, Katie and Bachhar, Ritesh and Islam, Tousif and Rifat, Nur E. M. and Gonzalez-Quesada, Kevin and Field, Scott E. and Khanna, Gaurav and Hughes, Scott A. and Varma, Vijay",
    title = "{Gravitational wave surrogate model for spinning, intermediate mass ratio binaries based on perturbation theory and numerical relativity}",
    eprint = "2407.18319",
    archivePrefix = "arXiv",
    primaryClass = "gr-qc",
    doi = "10.1103/PhysRevD.110.124069",
    journal = "Phys. Rev. D",
    volume = "110",
    number = "12",
    pages = "124069",
    year = "2024"
}

@article{Kaup:1968zz,
    author = "Kaup, David J.",
    title = "{Klein-Gordon Geon}",
    journal = "Phys. Rev.",
    volume = "172",
    pages = "1331-1342",
    year = "1968",
    doi = "10.1103/PhysRev.172.1331"
}

@article{Iacovelli:2022For,
    author = "Iacovelli, Francesco and Mancarella, Michele and Foffa, Stefano and Maggiore, Michele",
    title = "{Forecasting the Detection Capabilities of Third-generation Gravitational-wave Detectors Using GWFAST}",
    eprint = "2207.02771",
    archivePrefix = "arXiv",
    primaryClass = "gr-qc",
    doi = "10.3847/1538-4357/ac9cd4",
    journal = "Astrophys. J.",
    volume = "941",
    number = "2",
    pages = "208",
    year = "2022"
}

@article{Vallisneri:2007ev,
    author = "Vallisneri, Michele",
    title = "{Use and abuse of the Fisher information matrix in the assessment of gravitational-wave parameter-estimation prospects}",
    eprint = "gr-qc/0703086",
    archivePrefix = "arXiv",
    reportNumber = "LIGO-P070009-00-Z",
    doi = "10.1103/PhysRevD.77.042001",
    journal = "Phys. Rev. D",
    volume = "77",
    pages = "042001",
    year = "2008"
}

@article{Rodriguez:2013mla,
    author = "Rodriguez, Carl L. and Farr, Benjamin and Farr, Will M. and Mandel, Ilya",
    title = "{Inadequacies of the Fisher Information Matrix in gravitational-wave parameter estimation}",
    eprint = "1308.1397",
    archivePrefix = "arXiv",
    primaryClass = "astro-ph.IM",
    doi = "10.1103/PhysRevD.88.084013",
    journal = "Phys. Rev. D",
    volume = "88",
    number = "8",
    pages = "084013",
    year = "2013"
}

@article{Cutler:1994ys,
    author = "Cutler, Curt and Flanagan, Eanna E.",
    title = "{Gravitational waves from merging compact binaries: How accurately can one extract the binary's parameters from the inspiral wave form?}",
    eprint = "gr-qc/9402014",
    archivePrefix = "arXiv",
    reportNumber = "GRP-369",
    doi = "10.1103/PhysRevD.49.2658",
    journal = "Phys. Rev. D",
    volume = "49",
    pages = "2658--2697",
    year = "1994"
}

@article{deSouza:2023DALI,
    author = "de Souza, Josiel Mendon\c{c}a Soares and Sturani, Riccardo",
    title = "{GWDALI: A Fisher-matrix based software for gravitational wave parameter-estimation beyond Gaussian approximation}",
    eprint = "2307.10154",
    archivePrefix = "arXiv",
    primaryClass = "gr-qc",
    doi = "10.1016/j.ascom.2023.100759",
    journal = "Astron. Comput.",
    volume = "45",
    pages = "100759",
    year = "2023"
}

@article{Li:2021Tidofm,
    author = "Li, Yufeng and Heng, Ik Siong and Chan, Man Leong and Messenger, Chris and Fan, Xilong",
    title = "{Exploring the sky localization and early warning capabilities of third generation gravitational wave detectors in three-detector network configurations}",
    eprint = "2109.07389",
    archivePrefix = "arXiv",
    primaryClass = "astro-ph.IM",
    doi = "10.1103/PhysRevD.105.043010",
    journal = "Phys. Rev. D",
    volume = "105",
    number = "4",
    pages = "043010",
    year = "2022"
}

@article{Planck:2018vyg,
    author = "Aghanim, N. and others",
    collaboration = "Planck",
    title = "{Planck 2018 results. VI. Cosmological parameters}",
    eprint = "1807.06209",
    archivePrefix = "arXiv",
    primaryClass = "astro-ph.CO",
    doi = "10.1051/0004-6361/201833910",
    journal = "Astron. Astrophys.",
    volume = "641",
    pages = "A6",
    year = "2020",
    note = "[Erratum: Astron.Astrophys. 652, C4 (2021)]"
}

@article{Abac:2025BB,
    author = "Abac, Adrian and others",
    title = "{The Science of the Einstein Telescope}",
    eprint = "2503.12263",
    archivePrefix = "arXiv",
    primaryClass = "gr-qc",
    reportNumber = "ET-0036C-25",
    month = "3",
    year = "2025"
}

@article{Borhanian:2020GWB,
    author = "Borhanian, Ssohrab",
    title = "{GWBENCH: a novel Fisher information package for gravitational-wave benchmarking}",
    eprint = "2010.15202",
    archivePrefix = "arXiv",
    primaryClass = "gr-qc",
    doi = "10.1088/1361-6382/ac1618",
    journal = "Class. Quant. Grav.",
    volume = "38",
    number = "17",
    pages = "175014",
    year = "2021"
}

@article{Borhanian:2022czq,
    author = "Borhanian, Ssohrab and Sathyaprakash, B. S.",
    title = "{Listening to the Universe with next generation ground-based gravitational-wave detectors}",
    eprint = "2202.11048",
    archivePrefix = "arXiv",
    primaryClass = "gr-qc",
    doi = "10.1103/PhysRevD.110.083040",
    journal = "Phys. Rev. D",
    volume = "110",
    number = "8",
    pages = "083040",
    year = "2024"
}

@article{Akmal:1998cf,
    author = "Akmal, A. and Pandharipande, V. R. and Ravenhall, D. G.",
    title = "{The Equation of state of nucleon matter and neutron star structure}",
    eprint = "nucl-th/9804027",
    archivePrefix = "arXiv",
    doi = "10.1103/PhysRevC.58.1804",
    journal = "Phys. Rev. C",
    volume = "58",
    pages = "1804--1828",
    year = "1998"
}

@article{Finn:1992wt,
    author = "Finn, Lee S.",
    title = "{Detection, measurement and gravitational radiation}",
    eprint = "gr-qc/9209010",
    archivePrefix = "arXiv",
    reportNumber = "PRINT-93-0128 (NORTHWESTERN)",
    doi = "10.1103/PhysRevD.46.5236",
    journal = "Phys. Rev. D",
    volume = "46",
    pages = "5236--5249",
    year = "1992"
}

@article{Bodmer:1971we,
    author = "Bodmer, A. R.",
    title = "{Collapsed nuclei}",
    doi = "10.1103/PhysRevD.4.1601",
    journal = "Phys. Rev. D",
    volume = "4",
    pages = "1601--1606",
    year = "1971"
}

@article{Kleihaus:2011sx,
    author = "Kleihaus, Burkhard and Kunz, Jutta and Schneider, Stefanie",
    title = "{Stable Phases of Boson Stars}",
    eprint = "1109.5858",
    archivePrefix = "arXiv",
    primaryClass = "gr-qc",
    doi = "10.1103/PhysRevD.85.024045",
    journal = "Phys. Rev. D",
    volume = "85",
    pages = "024045",
    year = "2012"
}

@article{Chavanis:2011zi,
    author = "Chavanis, Pierre-Henri",
    title = "{Mass-radius relation of Newtonian self-gravitating Bose-Einstein condensates with short-range interactions: I. Analytical results}",
    eprint = "1103.2050",
    archivePrefix = "arXiv",
    primaryClass = "astro-ph.CO",
    doi = "10.1103/PhysRevD.84.043531",
    journal = "Phys. Rev. D",
    volume = "84",
    pages = "043531",
    year = "2011"
}

@article{Lattimer:2000nx,
    author = "Lattimer, J. M. and Prakash, M.",
    title = "{Neutron star structure and the equation of state}",
    eprint = "astro-ph/0002232",
    archivePrefix = "arXiv",
    doi = "10.1086/319702",
    journal = "Astrophys. J.",
    volume = "550",
    pages = "426",
    year = "2001"
}

@article{LIGOScientific:2018cki,
    author = "Abbott, B. P. and others",
    collaboration = "LIGO Scientific, Virgo",
    title = "{GW170817: Measurements of neutron star radii and equation of state}",
    eprint = "1805.11581",
    archivePrefix = "arXiv",
    primaryClass = "gr-qc",
    reportNumber = "LIGO-P1800115",
    doi = "10.1103/PhysRevLett.121.161101",
    journal = "Phys. Rev. Lett.",
    volume = "121",
    number = "16",
    pages = "161101",
    year = "2018"
}

@article{franciolini_how_2022,
    author = "Franciolini, Gabriele and Cotesta, Roberto and Loutrel, Nicholas and Berti, Emanuele and Pani, Paolo and Riotto, Antonio",
    title = "{How to assess the primordial origin of single gravitational-wave events with mass, spin, eccentricity, and deformability measurements}",
    eprint = "2112.10660",
    archivePrefix = "arXiv",
    primaryClass = "astro-ph.CO",
    reportNumber = "ET-0464A-21",
    doi = "10.1103/PhysRevD.105.063510",
    journal = "Phys. Rev. D",
    volume = "105",
    number = "6",
    pages = "063510",
    year = "2022"
}

@article{wang2021tidal,
  title={Tidal deformability of strange quark planets and strange dwarfs},
  author={Wang, Xu and Kuerban, Abudushataer and Geng, Jin-Jun and Xu, Fan and Zhang, Xiao-Li and Zuo, Bing-Jun and Yuan, Wen-Li and Huang, Yong-Feng},
  journal={Physical Review D},
  volume={104},
  number={12},
  pages={123028},
  year={2021},
  publisher={APS}
}

@article{Prakash:1995uw,
    author = "Prakash, M. and Cooke, J. R. and Lattimer, J. M.",
    title = "{Quark - hadron phase transition in protoneutron stars}",
    doi = "10.1103/PhysRevD.52.661",
    journal = "Phys. Rev. D",
    volume = "52",
    pages = "661--665",
    year = "1995"
}

@article{Reitze:2019iox,
    author = "Reitze, David and others",
    title = "{Cosmic Explorer: The U.S. Contribution to Gravitational-Wave Astronomy beyond LIGO}",
    eprint = "1907.04833",
    archivePrefix = "arXiv",
    primaryClass = "astro-ph.IM",
    reportNumber = "LIGO-P1900316",
    journal = "Bull. Am. Astron. Soc.",
    volume = "51",
    number = "7",
    pages = "035",
    year = "2019"
}

@article{Chatziioannou:2020pqz,
    author = "Chatziioannou, Katerina",
    title = "{Neutron star tidal deformability and equation of state constraints}",
    eprint = "2006.03168",
    archivePrefix = "arXiv",
    primaryClass = "gr-qc",
    doi = "10.1007/s10714-020-02754-3",
    journal = "Gen. Rel. Grav.",
    volume = "52",
    number = "11",
    pages = "109",
    year = "2020"
}

@article{Gurlebeck:2015xpa,
    author = {G{\"u}rlebeck, Norman},
    title = "{No-hair theorem for Black Holes in Astrophysical Environments}",
    eprint = "1503.03240",
    archivePrefix = "arXiv",
    primaryClass = "gr-qc",
    doi = "10.1103/PhysRevLett.114.151102",
    journal = "Phys. Rev. Lett.",
    volume = "114",
    number = "15",
    pages = "151102",
    year = "2015"
}

@article{Damour:2012yf,
    author = "Damour, Thibault and Nagar, Alessandro and Villain, Loic",
    title = "{Measurability of the tidal polarizability of neutron stars in late-inspiral gravitational-wave signals}",
    eprint = "1203.4352",
    archivePrefix = "arXiv",
    primaryClass = "gr-qc",
    doi = "10.1103/PhysRevD.85.123007",
    journal = "Phys. Rev. D",
    volume = "85",
    pages = "123007",
    year = "2012"
}

@article{LIGOScientific:2021sio,
    author = "Abbott, R. and others",
    collaboration = "LIGO Scientific, VIRGO, KAGRA",
    title = "{Tests of General Relativity with GWTC-3}",
    eprint = "2112.06861",
    archivePrefix = "arXiv",
    primaryClass = "gr-qc",
    reportNumber = "LIGO-P2100275",
    month = "12",
    year = "2021"
}

@article{KAGRA:2021vkt,
    author = "Abbott, R. and others",
    collaboration = "KAGRA, VIRGO, LIGO Scientific",
    title = "{GWTC-3: Compact Binary Coalescences Observed by LIGO and Virgo during the Second Part of the Third Observing Run}",
    eprint = "2111.03606",
    archivePrefix = "arXiv",
    primaryClass = "gr-qc",
    reportNumber = "LIGO-P2000318",
    doi = "10.1103/PhysRevX.13.041039",
    journal = "Phys. Rev. X",
    volume = "13",
    number = "4",
    pages = "041039",
    year = "2023"
}

@article{evstafyeva_gravitational-wave_2024,
	title = {Gravitational-{Wave} {Data} {Analysis} with {High}-{Precision} {Numerical} {Relativity} {Simulations} of {Boson} {Star} mergers},
	volume = {133},
	issn = {0031-9007, 1079-7114},
	url = {http://arxiv.org/abs/2406.02715},
	doi = {10.1103/PhysRevLett.133.131401},
	number = {13},
	urldate = {2025-01-05},
	journal = {Physical Review Letters},
	author = {Evstafyeva, Tamara and Sperhake, Ulrich and Romero-Shaw, Isobel and Agathos, Michalis},
	month = sep,
	year = {2024},
	note = {arXiv:2406.02715 [gr-qc]},
	pages = {131401},
}

@article{pacilio_ranking_2022,
    author = "Pacilio, Costantino and Maselli, Andrea and Fasano, Margherita and Pani, Paolo",
    title = "{Ranking Love Numbers for the Neutron Star Equation of State: The Need for Third-Generation Detectors}",
    eprint = "2104.10035",
    archivePrefix = "arXiv",
    primaryClass = "gr-qc",
    doi = "10.1103/PhysRevLett.128.101101",
    journal = "Phys. Rev. Lett.",
    volume = "128",
    number = "10",
    pages = "101101",
    year = "2022"
}

@article{Crescimbeni:2024qrq,
    author = "Crescimbeni, Francesco and Franciolini, Gabriele and Pani, Paolo and Vaglio, Massimo",
    title = "{Cosmology and nuclear physics implications of a subsolar gravitational-wave event}",
    eprint = "2408.14287",
    archivePrefix = "arXiv",
    primaryClass = "astro-ph.HE",
    reportNumber = "CERN-TH-2024-121",
    doi = "10.1103/PhysRevD.111.083538",
    journal = "Phys. Rev. D",
    volume = "111",
    number = "8",
    pages = "083538",
    year = "2025"
}

@article{DeLuca:2019buf,
    author = "De Luca, Valerio and Franciolini, Gabriele and Pani, Paolo and Riotto, Antonio",
    title = "{The Evolution of Primordial Black Holes and their Spins}",
    eprint = "1909.03883",
    archivePrefix = "arXiv",
    primaryClass = "astro-ph.CO",
    journal = "JCAP",
    volume = "04",
    pages = "052",
    year = "2020",
    doi = "10.1088/1475-7516/2020/04/052"
}

@article{Green:2020jor,
    author = "Green, Anne M. and Kavanagh, Bradley J.",
    title = "{Primordial Black Holes as a dark matter candidate}",
    eprint = "2007.10722",
    archivePrefix = "arXiv",
    primaryClass = "astro-ph.CO",
    doi = "10.1088/1361-6471/abc534",
    journal = "J. Phys. G",
    volume = "48",
    number = "4",
    pages = "043001",
    year = "2021"
}

@article{Carr:2023tpt,
    author = "Carr, Bernard and Green, Anne M. and Kuhnel, Florian",
    title = "{Primordial Black Holes as a Probe of Cosmology and High Energy Physics}",
    eprint = "2306.03903",
    archivePrefix = "arXiv",
    primaryClass = "astro-ph.CO",
    journal = "Phys. Rept.",
    volume = "1054",
    pages = "1--67",
    year = "2024",
    doi = "10.1016/j.physrep.2024.01.002"
}

@article{Hawking:1971ei,
    author = "Hawking, S.",
    title = "{Gravitationally collapsed objects of very low mass}",
    journal = "Mon. Not. Roy. Astron. Soc.",
    volume = "152",
    pages = "75",
    year = "1971"
}

@article{Carr:1974nx,
    author = "Carr, B. J. and Hawking, S. W.",
    title = "{Black holes in the early Universe}",
    journal = "Mon. Not. Roy. Astron. Soc.",
    volume = "168",
    pages = "399--415",
    year = "1974"
}

@article{Nitz:2022ltl,
    author = "Nitz, Alexander H. and Wang, Yi-Fan",
    title = "{Broad search for gravitational waves from subsolar-mass binaries through LIGO and Virgo{\textquoteright}s third observing run}",
    eprint = "2202.11024",
    archivePrefix = "arXiv",
    primaryClass = "astro-ph.HE",
    doi = "10.1103/PhysRevD.106.023024",
    journal = "Phys. Rev. D",
    volume = "106",
    number = "2",
    pages = "023024",
    year = "2022"
}

@article{LVK:2022ydq,
    author = "Abbott, R. and others",
    collaboration = "LVK",
    title = "{Search for subsolar-mass black hole binaries in the second part of Advanced LIGO{\textquoteright}s and Advanced Virgo{\textquoteright}s third observing run}",
    eprint = "2212.01477",
    archivePrefix = "arXiv",
    primaryClass = "astro-ph.HE",
    doi = "10.1093/mnras/stad588",
    journal = "Mon. Not. Roy. Astron. Soc.",
    volume = "524",
    number = "4",
    pages = "5984--5992",
    year = "2023",
    note = "[Erratum: Mon.Not.Roy.Astron.Soc. 526, 6234 (2023)]"
}

@article{Wang:2016ana,
    author = "Wang, Sai and Wang, Yi-Fan and Huang, Qing-Guo and Li, Tjonnie G. F.",
    title = "{Constraints on the Primordial Black Hole Abundance from the First Advanced LIGO Observation Run Using the Stochastic Gravitational-Wave Background}",
    eprint = "1610.08725",
    archivePrefix = "arXiv",
    primaryClass = "astro-ph.CO",
    doi = "10.1103/PhysRevLett.120.191102",
    journal = "Phys. Rev. Lett.",
    volume = "120",
    number = "19",
    pages = "191102",
    year = "2018"
}

@article{Giudice:2016zpa,
    author = "Giudice, Gian F. and McCullough, Matthew and Urbano, Alfredo",
    title = "{Hunting for Dark Particles with Gravitational Waves}",
    eprint = "1605.01209",
    archivePrefix = "arXiv",
    primaryClass = "hep-ph",
    reportNumber = "CERN-TH-2016-106",
    doi = "10.1088/1475-7516/2016/10/001",
    journal = "JCAP",
    volume = "10",
    pages = "001",
    year = "2016"
}

@article{Carr:2016drx,
    author = "Carr, Bernard and Kuhnel, Florian and Sandstad, Lennart",
    title = "{Primordial Black Holes as Dark Matter}",
    eprint = "1607.06077",
    archivePrefix = "arXiv",
    primaryClass = "astro-ph.CO",
    journal = "Phys. Rev. D",
    volume = "94",
    pages = "083504",
    year = "2016",
    doi = "10.1103/PhysRevD.94.083504"
}

@article{Suwa:2018yxd,
    author = "Suwa, Yudai and Yoshida, Takashi and Shibata, Masaru and Umeda, Hideyuki and Takahashi, Koh",
    title = "{On the minimum mass of neutron stars}",
    eprint = "1808.02368",
    archivePrefix = "arXiv",
    primaryClass = "astro-ph.HE",
    journal = "Mon. Not. Roy. Astron. Soc.",
    volume = "481",
    pages = "3305",
    year = "2018",
    doi = "10.1093/mnras/sty2460"
}

@article{Kacanja:2025,
    author = "Kacanja, Keisi and Nitz, Alexander H.",
    title = "{A Search for Low-mass Neutron Stars in the Third Observing Run of Advanced LIGO and Virgo}",
    eprint = "2412.05369",
    archivePrefix = "arXiv",
    primaryClass = "astro-ph.HE",
    doi = "10.3847/1538-4357/adc9a5",
    journal = "Astrophys. J.",
    volume = "984",
    number = "1",
    pages = "61",
    year = "2025"
}

@article{Binnington:2009bb,
    author = "Binnington, Taylor and Poisson, Eric",
    title = "{Relativistic theory of tidal Love numbers}",
    eprint = "0906.1366",
    archivePrefix = "arXiv",
    primaryClass = "gr-qc",
    journal = "Phys. Rev. D",
    volume = "80",
    pages = "084018",
    year = "2009",
    doi = "10.1103/PhysRevD.80.084018"
}

@article{Damour:2009vw,
    author = "Damour, Thibault and Nagar, Alessandro",
    title = "{Relativistic tidal properties of neutron stars}",
    eprint = "0906.0096",
    archivePrefix = "arXiv",
    primaryClass = "gr-qc",
    journal = "Phys. Rev. D",
    volume = "80",
    pages = "084035",
    year = "2009",
    doi = "10.1103/PhysRevD.80.084035"
}

@article{Poisson:2014cga,
    author = "Poisson, Eric",
    title = "{Tidal deformation of a slowly rotating black hole}",
    eprint = "1404.2192",
    archivePrefix = "arXiv",
    primaryClass = "gr-qc",
    journal = "Phys. Rev. D",
    volume = "91",
    pages = "044004",
    year = "2015",
    doi = "10.1103/PhysRevD.91.044004"
}

@article{Begnoni:2025mtz,
    author = "Begnoni, Andrea and Del Pozzo, Walter and Pegorin, Matteo and Pomper, Joachim and Ricciardone, Angelo",
    title = "{Tests of General Relativity with Einstein Telescope}",
    eprint = "2511.07520",
    archivePrefix = "arXiv",
    primaryClass = "gr-qc",
    month = "11",
    year = "2025"
}

@article{Dupletsa:2024gfl,
    author = "Dupletsa, Ulyana and Harms, Jan and Ng, Ken K. Y. and Tissino, Jacopo and Santoliquido, Filippo and Cozzumbo, Andrea",
    title = "{Validating prior-informed Fisher-matrix analyses against GWTC data}",
    eprint = "2404.16103",
    archivePrefix = "arXiv",
    primaryClass = "gr-qc",
    doi = "10.1103/PhysRevD.111.024036",
    journal = "Phys. Rev. D",
    volume = "111",
    number = "2",
    pages = "024036",
    year = "2025"
}

@book{maggiore2008gravitational,
  title={Gravitational waves},
  author={Maggiore, Michele},
  volume={2},
  year={2008},
  publisher={Oxford university press}
}

@article{Crescimbeni:2024tidal,
    author = "Crescimbeni, Francesco and Franciolini, Gabriele and Pani, Paolo and Riotto, Antonio",
    title = "{Can we identify primordial black holes? Tidal tests for subsolar-mass gravitational-wave observations}",
    eprint = "2402.18656",
    archivePrefix = "arXiv",
    primaryClass = "astro-ph.HE",
    reportNumber = "CERN-TH-2024-026",
    doi = "10.1103/PhysRevD.109.124063",
    journal = "Phys. Rev. D",
    volume = "109",
    number = "12",
    pages = "124063",
    year = "2024"
}

@article{Sennett:2017etc,
    author = "Sennett, Noah and Hinderer, Tanja and Steinhoff, Jan and Buonanno, Alessandra and Ossokine, Sergey",
    title = "{Distinguishing Boson Stars from Black Holes and Neutron Stars from Tidal Interactions in Inspiraling Binary Systems}",
    eprint = "1704.08651",
    archivePrefix = "arXiv",
    primaryClass = "gr-qc",
    journal = "Phys. Rev. D",
    volume = "96",
    pages = "024002",
    year = "2017",
    doi = "10.1103/PhysRevD.96.024002"
}

@article{Harada:2017fjm,
    author = "Harada, Tomohiro and Yoo, Chul-Moon and Kohri, Kazunori and Nakao, Ken-Ichi",
    title = "{Spins of primordial black holes formed in the matter-dominated phase of the Universe}",
    eprint = "1707.03595",
    archivePrefix = "arXiv",
    primaryClass = "gr-qc",
    journal = "Phys. Rev. D",
    volume = "96",
    pages = "083517",
    year = "2017",
    doi = "10.1103/PhysRevD.96.083517"
}

@article{Chiba:2017rvs,
    author = "Chiba, Takeshi and Yokoyama, Shuichiro",
    title = "{Spin Distribution of Primordial Black Holes}",
    eprint = "1704.06573",
    archivePrefix = "arXiv",
    primaryClass = "gr-qc",
    journal = "PTEP",
    volume = "2017",
    number = "8",
    pages = "083E01",
    year = "2017",
    doi = "10.1093/ptep/ptx087"
}

@article{Ruffini:1969qy,
  author = "Ruffini, Remo and Bonazzola, Silvano",
  title = "{Systems of selfgravitating particles in general relativity and the concept of an equation of state}",
  journal = "Phys. Rev.",
  volume = "187",
  pages = "1767--1783",
  year = "1969",
  doi = "10.1103/PhysRev.187.1767"
}

@article{Hinderer:2007mb,
  author = "Hinderer, Tanja",
  title = "{Tidal Love numbers of neutron stars}",
  journal = "Astrophys. J.",
  volume = "677",
  pages = "1216--1220",
  year = "2008",
  eprint = "0711.2420",
  archivePrefix = "arXiv",
  primaryClass = "astro-ph",
  doi = "10.1086/533487"
}

@article{Abbott:2018exr,
  author = "Abbott, B. P. and others",
  collaboration = "LIGO Scientific, Virgo",
  title = "{GW170817: Measurements of neutron star radii and equation of state}",
  journal = "Phys. Rev. Lett.",
  volume = "121",
  number = "16",
  pages = "161101",
  year = "2018",
  eprint = "1805.11581",
  archivePrefix = "arXiv",
  primaryClass = "gr-qc",
  doi = "10.1103/PhysRevLett.121.161101"
}

@article{Evans:2021gyd,
    author = "Evans, Matthew and others",
    title = "{A Horizon Study for Cosmic Explorer: Science, Observatories, and Community}",
    eprint = "2109.09882",
    archivePrefix = "arXiv",
    primaryClass = "astro-ph.IM",
    year = "2021"
}

@article{Punturo:2010zz,
    author = "Punturo, M. and others",
    title = "{The Einstein Telescope: A third-generation gravitational wave observatory}",
    journal = "Class. Quant. Grav.",
    volume = "27",
    pages = "194002",
    year = "2010",
    doi = "10.1088/0264-9381/27/19/194002"
}

@article{Stasenko:2024pbh,
    author = "Stasenko, Viktor",
    title = "{Redshift evolution of primordial black hole merger rate}",
    eprint = "2403.11325",
    archivePrefix = "arXiv",
    primaryClass = "astro-ph.CO",
    doi = "10.1103/PhysRevD.109.123546",
    journal = "Phys. Rev. D",
    volume = "109",
    number = "12",
    pages = "123546",
    year = "2024"
}

@article{Poulin:2017pbh,
  author       = {Poulin, Vivian and Serpico, Pasquale D. and Calore, Francesca and Clesse, Sebastien and Kohri, Kazunori},
  title        = {CMB bounds on dark matter including ${\rm PBH}$s as a candidate},
  journal      = {Phys. Rev. D},
  volume       = {96},
  pages        = {083524},
  year         = {2017},
  doi          = {10.1103/PhysRevD.96.083524},
  eprint       = {1707.04206},
  archivePrefix= {arXiv},
  primaryClass = {astro-ph.CO}
}

@article{Serpico:2020pbh,
  author       = {Serpico, Pasquale D. and Poulin, Vivian and Inoue, Yoshiaki and Kohri, Kazunori},
  title        = {Probing PBH dark matter with radio and X-ray observations},
  journal      = {Phys. Rev. Res.},
  volume       = {2},
  pages        = {023204},
  year         = {2020},
  doi          = {10.1103/PhysRevResearch.2.023204},
  eprint       = {2002.10771},
  archivePrefix= {arXiv},
  primaryClass = {astro-ph.CO}
}

@article{Sicilia:2021gtu,
    author = "Sicilia, Alex and Lapi, Andrea and Boco, Lumen and Spera, Mario and Di Carlo, Ugo N. and Mapelli, Michela and Shankar, Francesco and Alexander, David M. and Bressan, Alessandro and Danese, Luigi",
    title = "{The Black Hole Mass Function Across Cosmic Times. I. Stellar Black Holes and Light Seed Distribution}",
    eprint = "2110.15607",
    archivePrefix = "arXiv",
    primaryClass = "astro-ph.GA",
    doi = "10.3847/1538-4357/ac34fb",
    journal = "Astrophys. J.",
    volume = "924",
    number = "2",
    pages = "56",
    year = "2022"
}

@article{Gamba:2020ljo,
  author       = {Gamba, Rossella and Breschi, Matteo and Bernuzzi, Sebastiano and Agathos, Michalis and Nagar, Alessandro},
  title        = {Waveform systematics in the gravitational‐wave inference of tidal parameters and equation of state from binary neutron star signals},
  journal      = {Phys.\ Rev.\ D},
  volume       = {103},
  number       = {12},
  pages        = {124015},
  year         = {2021},
  doi          = {10.1103/PhysRevD.103.124015},
  eprint       = {arXiv:2009.08467},
  archivePrefix= {arXiv},
  primaryClass = {gr-qc},
}

\end{document}